\documentclass[a4paper,fleqn]{cas-dc}

\usepackage{graphicx}
\usepackage{subcaption}
\usepackage{placeins}
\usepackage{import}
\usepackage{pgfplots}
\pgfplotsset{compat=newest}
\usepackage[]{standalone}

\usepackage{cleveref}
\crefname{figure}{Fig.}{Figs.}
\crefname{table}{Table}{Tables}
\crefname{equation}{Eq.}{Eqs.}

\usepackage{cancel}
\usepackage[percent]{overpic}

\usepackage[numbers]{natbib}

\def\tsc#1{\csdef{#1}{\textsc{\lowercase{#1}}\xspace}}
\tsc{WGM}
\tsc{QE}
\tsc{EP}
\tsc{PMS}
\tsc{BEC}
\tsc{DE}

\begin{document}
\let\WriteBookmarks\relax
\def\floatpagepagefraction{1}
\def\textpagefraction{.001}

\shorttitle{TPMS evacuation}

\shortauthors{Aashish K Gupta et~al.}

\title [mode = title]{Effect of cohesion on the gravity-driven evacuation of metal powder through Triply-Periodic Minimal Surface structures}                      


\author[1]{Aashish K Gupta}
\ead{aashish.gupta@ed.ac.uk}
\credit{Conceptualization of this study, Methodology, Data curation, Analysis, Writing - Original draft preparation}
\cormark[1]

\author[1]{Christopher Ness}
\credit{Methodology, Analysis, Writing - Original draft preparation}

\author[1,2]{Sina Haeri}
\credit{Conceptualization of this study, Analysis, Editing, Project supervision}

\affiliation[1]{School of Engineering, University of Edinburgh, Edinburgh EH9 3FG, United Kingdom}
\affiliation[2]{HR Wallingford, Howbery Park, Wallingford, Oxfordshire, OX10 8BA, United Kingdom}

\cortext[cor1]{Corresponding author}


\begin{abstract}
Evacuating the powder trapped inside the complex cavities of Triply Periodic Minimal Surface (TPMS) structures remains a major challenge in metal-powder-based additive manufacturing.  
The Discrete Element Method offers valuable insights into this evacuation process, enabling the design of effective de-powdering strategies. In this study, we simulate gravity-driven evacuation of trapped powders from inside unit cells of various TPMS structures. We systematically investigate the role of cohesive energy density in shaping the discharge profile. 
Overall, we conclude that the Schwarz-P and Gyroid topologies enable the most efficient powder evacuation, remaining resilient to cohesion-induced flow hindrance. Furthermore, for the two  unit cells, we analyse detailed kinematics and interpret the results in relation to particle overlaps and contact force distributions.
\end{abstract}




\maketitle

\section{Introduction}
\begin{figure*}
  \centering           
  \begin{subfigure}{0.25\textwidth}
    \centering
    \includegraphics[width=\linewidth]{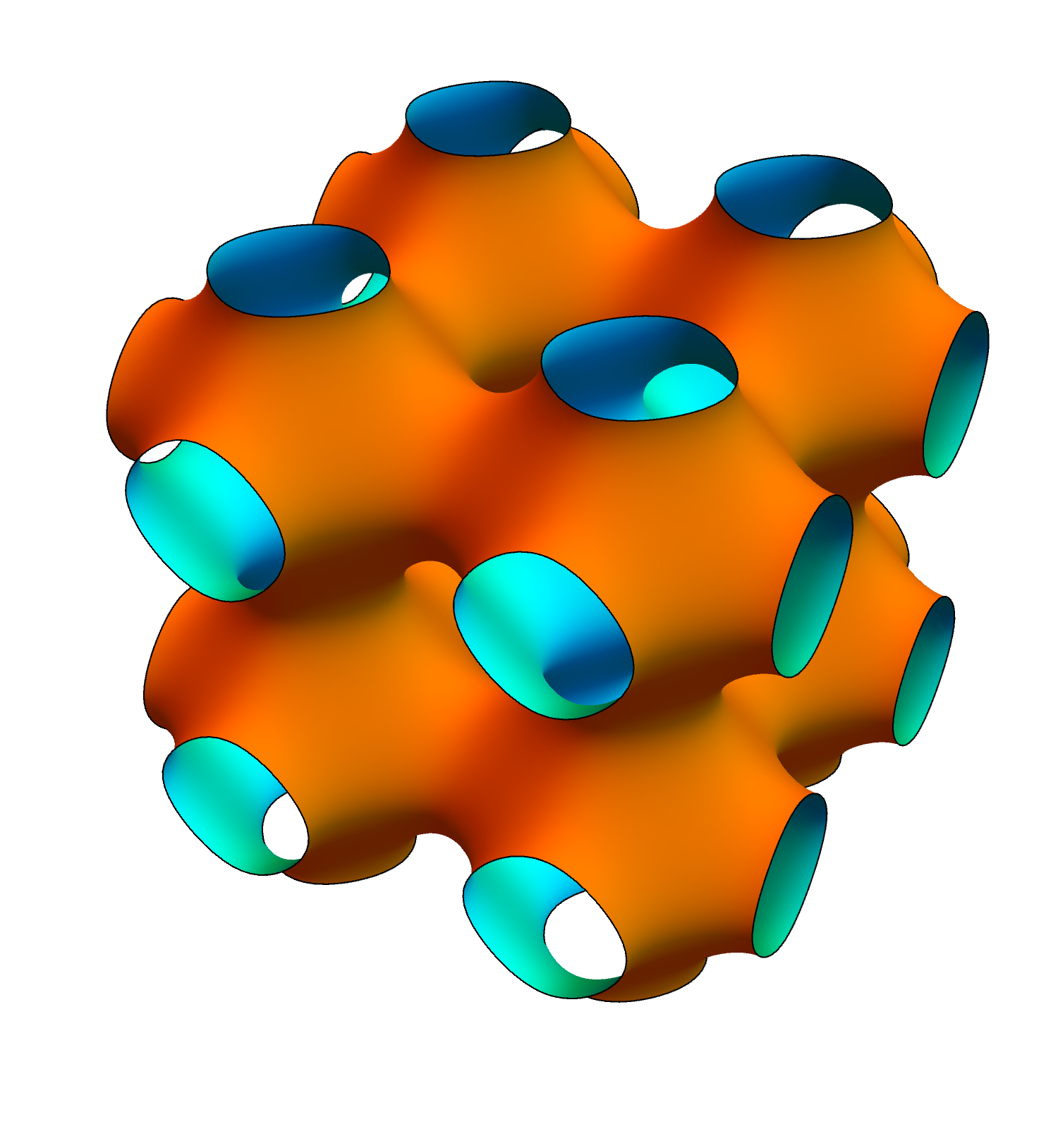}
    \caption{}
  \end{subfigure}           
  \begin{subfigure}{0.25\textwidth}
    \centering
    \includegraphics[width=\linewidth]{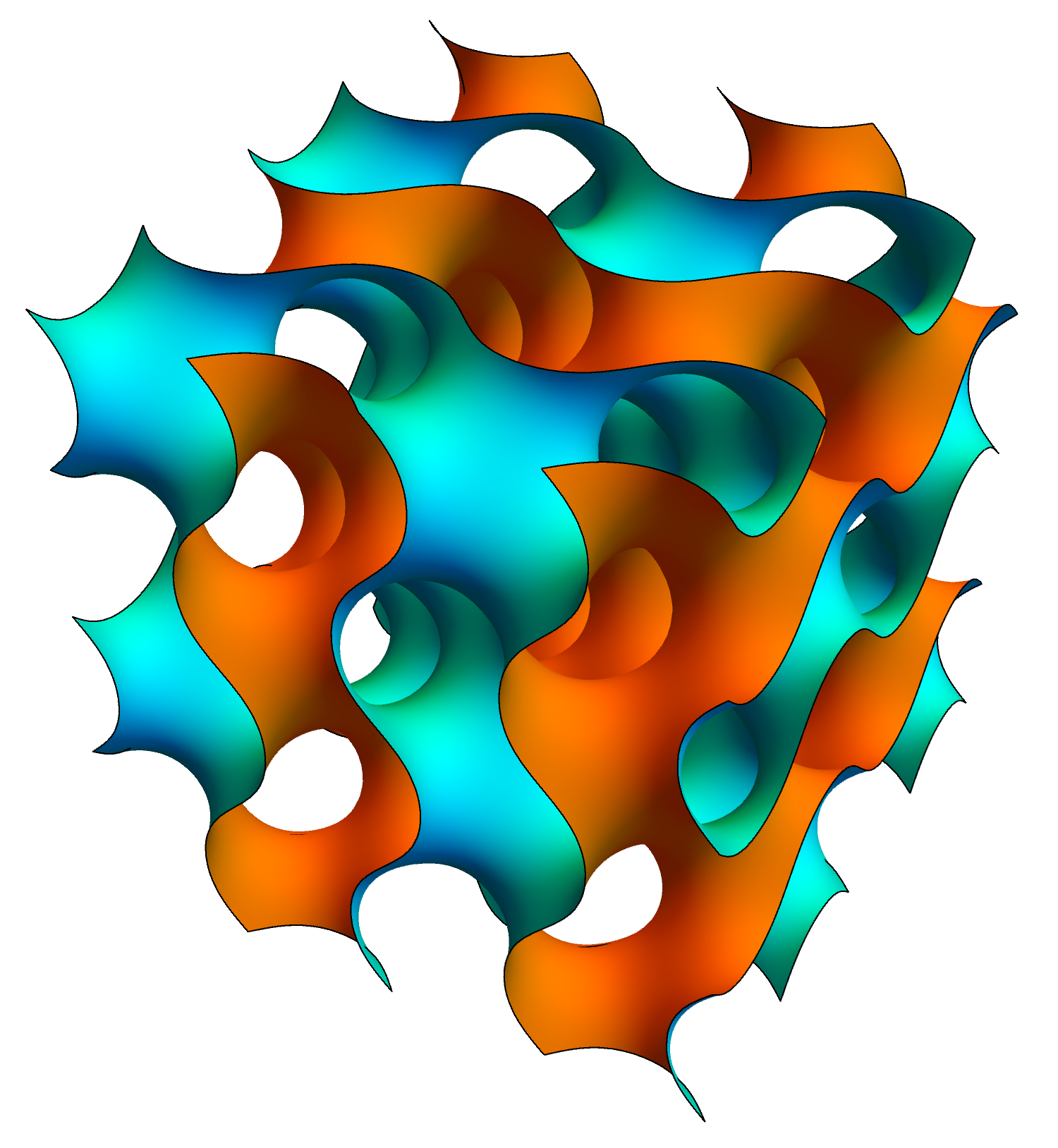}
    \caption{}
  \end{subfigure}
  \begin{subfigure}{0.25\textwidth}
    \centering
    \includegraphics[width=\linewidth]{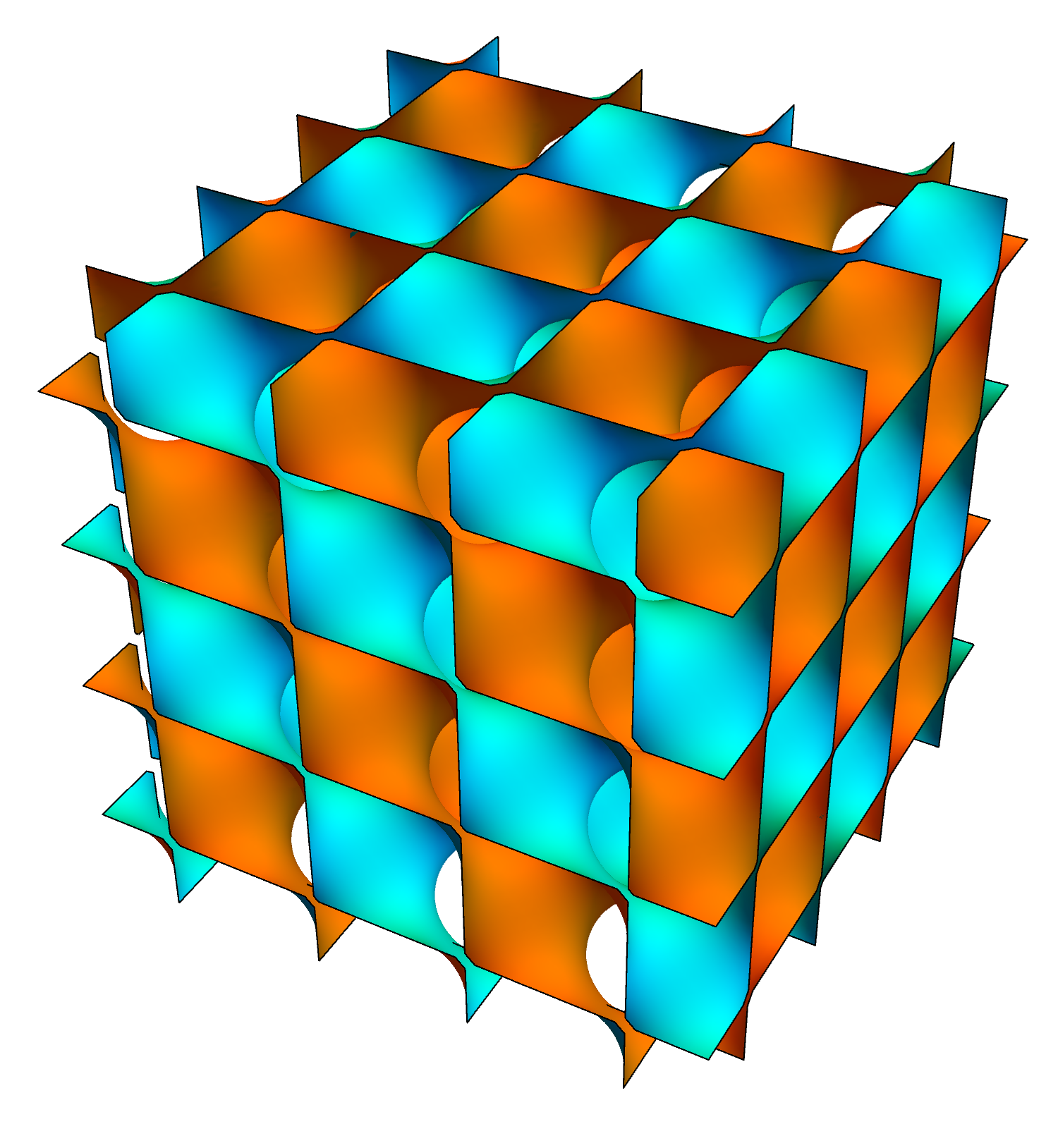}
    \caption{}
  \end{subfigure}

  \caption{$2\times2\times2$ array-structures built from TPMS unit cells commonly employed in Additive Manufacturing (AM). Panel (a) is based on a Schwarz-P unit cell, (b) on a Gyroid, and (c) on a Diamond unit cell. The two disjoint regions in each of the structures are shown in orange and cyan.}
  \label{TPMS array}
\end{figure*}

Amongst all the surfaces satisfying a given set of constraints, the one that has the least surface area is known as a minimal surface. An example of such a surface is the soap film formed by dipping a wireframe into a soap solution. A minimal surface whose repetition along three linearly independent directions generates a set of intertwined but non-intersecting structures belongs to the family of Triply Periodic Minimal Surfaces (TPMS). In nature, these surfaces manifest themselves in a variety of places such as biological membranes, wings of certain butterflies (rendering them specific colours), the skeletal plates of sea urchins, and the microstructure of nanoporous gold \cite{al2019multifunctional}. Inspired by these natural porous architectures, with mathematically controllable geometric features, TPMS scaffolds are being explored for revolutionary applications such as bone tissue regeneration \cite{dong2021application}, lightweight impact-resistant structure fabrication \citep{yu2019investigation}, and high-performance compact heat exchanger design in miniaturized electronics~\cite{alteneiji2022heat}. Due to the complex topology and intricate internal pores, conventional cutting or milling methods (that is, subtractive manufacturing) cannot be applied to fabricate TPMS structures, and the use of Additive Manufacturing (AM) is necessary \cite{feng2022triply}. \Cref{TPMS array} shows $2 \times 2 \times 2$ arrays of the commonly used TPMS structures in AM. The structures partition the entire space into two disjointed regions (shown in orange and the cyan) which can be seamlessly extended in the preferred direction(s).

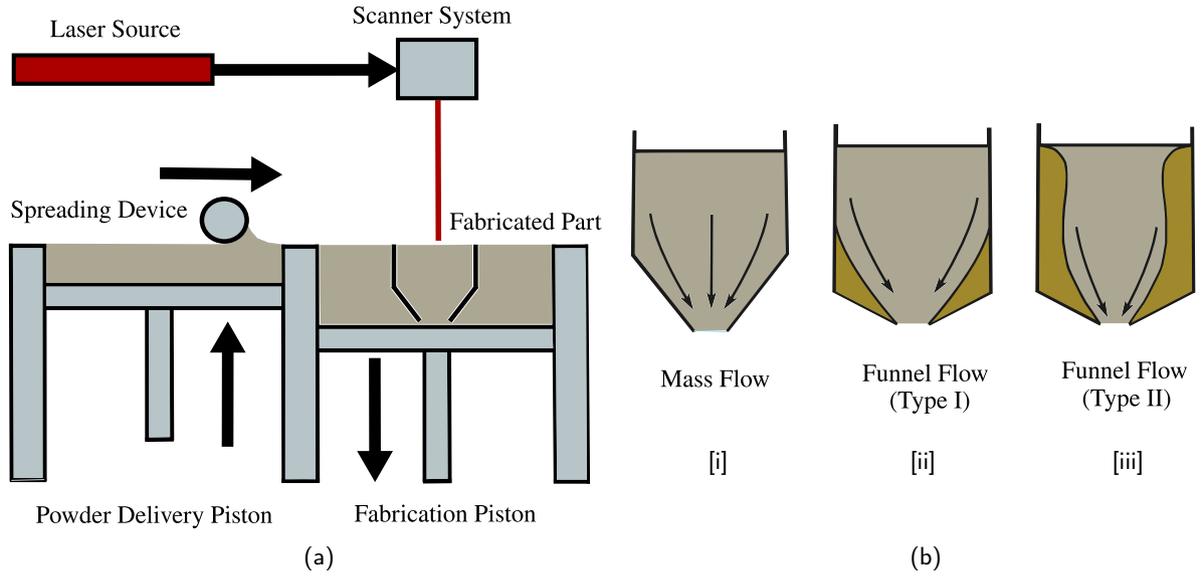
\begin{figure*}
    \centering
    \begin{overpic}[width=0.9\textwidth]{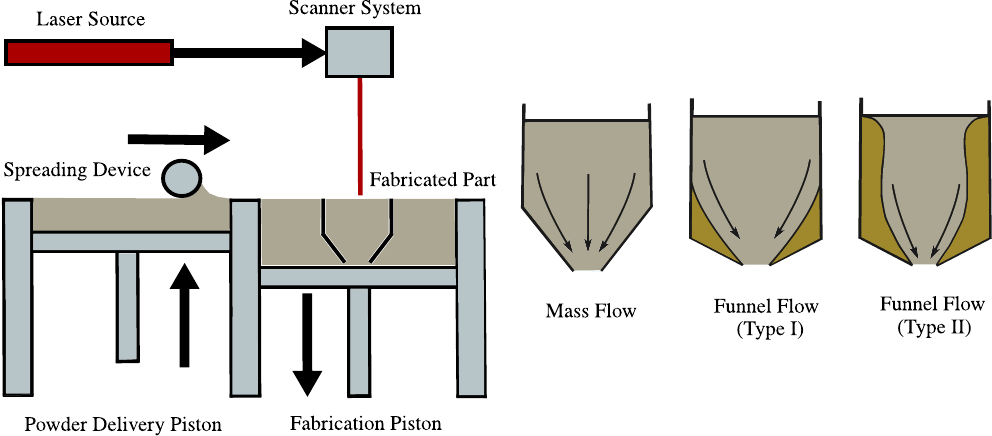}
     \put(25,-3) {(a)}
     \put(76,-3) {(b)}
     \put(59,5){[i]}
     \put(76,5){[ii]}
     \put(93,5){[iii]}
     \end{overpic}
     \vspace{1.5em}
      \caption{(a) Schematic of the Powder Bed Fusion-based AM of a bunker, and (b) the flow regimes encountered during the evacuation of the unused trapped powder. (b)[i] shows mass flow, where the entire granular medium is in motion, (b)[ii] and b[iii] show funnel flow, where there's a slow-moving/ stagnated region of material close to the bunker with a faster moving central core, with the passive region being localised at the outlet in the former, and spread over the entire system in the latter.}
    \label{AM_merged}
\end{figure*}

A typical metal powder-based AM process begins with the creation of a 3D CAD model, which is then sliced into several discrete layers. The pattern on each layer is recreated by selectively melting a bed of fine powder with a high-intensity laser beam (as shown in  \cref{AM_merged}). Once the replication of a layer is over, a new layer of powder is spread on the existing bed, and the process continues until all the sections of the CAD model have been mapped. After the completion of the selective melting stage, the powder trapped inside the cavities of the final build needs to be evacuated. However, a problem arises when the dimension of the pores in the structure to be produced is within 1-2 orders of magnitude of the diameter of powder grains \citep{zuriguel2005jamming}. This makes it challenging to empty the powder trapped within the pores after the component has been formed.

There is a large literature on the flow of granular materials under gravity emanating mainly from decades of research in the context of silos— identified as bins, hoppers, and bunkers depending on their shapes, which are vessels employed for storing and discharging of particulate solids in industrial settings. Experiments show that the mass flow rate of these materials through an orifice is independent of the head or height of material above the exit slot, provided the head is larger than a few multiples of the size of the exit slot \citep{rao2008introduction}. \citet{beverloo1961flow} proposed an empirical relation for the mass discharge rate of granular solids, based on their experimental investigation of a number of seeds, which was later generalized to 
\begin{equation}
    Q=C\rho_b \sqrt{g} \, (D-kd_p)^{5/2}\text{,}
    \label{beverloo_eqn}
\end{equation}
where C and k are constants depending on the hopper or silo geometry and particle properties, and $\rho_b$ is the bulk density of the material. The correlation has validity when the orifice diameter is sufficiently larger than the particle diameter ($D>>d_p$) such that there are no permanent flow interruptions due to arch formation at the outlet. The discharge law was modified with an exponential multiplicative factor, and the parameter $k$ was dropped altogether to reproduce a wide range of experimental measurements,
especially to fit those at small orifice sizes by \citet{mankoc2007flow} as
\begin{equation}
    Q=C'\rho_b \sqrt{g} \,[1-\frac{1}{2}e^{-b(D-d_p)}] \, (D-d_p)^{5/2}\text{,}
\end{equation}
with $C'$ and $b$ being fitting parameters. The corrective factor was suggested to be possibly related to the apparent density near the outlet of the silo.

As far as the flow within the silos is concerned, it can either be mass flow type or funnel flow type depending upon the steepness of the hopper section, and the intrinsic flowability of the material (\cref{AM_merged}). In mass flow, the entire material inside the silo is in motion, while in funnel flow there is a rapidly moving central core surrounded by shoulders of stagnant or creeping material \citep{rao2008introduction}. Based on the premise of a funnel-type flow, \citet{oldal2012outflow} derived an expression similar to \cref{beverloo_eqn}  speculating that the constant discharge rate of silos could be a consequence of the formation and collapse of arches in the bulk before the exit. The conjecture was based on the fact that the discharge rate, as reported from several preceding experiments, wasn't absolutely steady, but rather kept pulsating about a mean. In their paper, they assume the transient arch is a parabolic surface ending at the edge of the opening, and is described by 
\begin{equation}
   f(x)=h\left(1-\left(\frac{2x}{D}\right)^2\right) \text{.}
   \label{parabola}
\end{equation}
Here $f(x)$ is the height of a point on the arch measured from the plane of the exit opening at a distance $x$ from the centre of the opening (of diameter $D$), and $h$ is the maximum height of the arch, which occurs at the centre. A particle exiting the system is under free fall from this surface, where its velocity was zero; after having fallen through the distance of $f(x)$, it acquires a velocity of $\sqrt{2gf(x)}$, making the overall mass discharge rate at the orifice $\propto g^{1/2} d^{5/2}$. The form of their proposed velocity profile at the exit was also corroborated by their own experiments. 
\citet{janda2012flow} report similar velocity profiles for a wide range of apertures from their 
high-speed photography of particles at the exit. In fact, $\sqrt{gD}$ was found to be a good approximation for the velocity at the centre of the exit, making the height of the arch ($h$) in \cref{parabola} be $D/2$. Additionally, they also looked into the solid fraction profile close to the exit and found it $\propto [1-\alpha_1 e^{-D/2\alpha_2}]$, in agreement with the reasoning of \cite{mankoc2007flow}.
    
As the orifice size drops to the same order of magnitude as the particle diameter, the propensity of the outlet to clog increases.
\citet{geiger2007granular} in their empirical investigation of the jamming onset point defined as the ratio of the hopper opening $(D)$ to the mean particle diameter($d_p$) when the flow arrests—of relatively free-flowing granules. For spherical beads, the dimensionless critical diameter $(D_c/d_p)$ is $4.94 \pm 0.03$. 

Despite extensive studies on silos, discharge in more complex geometries has received limited attention. The role of cohesion in such systems has also not been systematically explored. Stronger cohesion is anticipated to increase the critical orifice size for jamming, and transient arching behaviours may still leave signatures in the flow dynamics. In metal powder-based AM, cohesive effects—stemming from van der Waals interactions, humidity, and electrostatic attraction—can significantly influence powder spreadability and flowability \cite{khajepor2023effects, haeri2017discrete, haeri2017optimisation, haeri2018impact, haeri2020effects}. Previous studies on powder-layer spreading in AM have examined Bond numbers (a dimensionless measure of cohesion relative to gravitational forces) up to approximately 300–400 (for Ti-6Al-4V powder)\cite{roy2024structural, he2020linking}. From the perspective of achieving optimal surface finish in the final component, finer powders are preferred. However, this preference exacerbates cohesion-related challenges, as smaller particles tend to have higher cohesion, complicating the de-powdering process.

In this article, we first develop a protocol to generate complex TPMS geometries packed with particles. We then explore how cohesion influences powder discharge profile in commonly used TPMS unit cells: Schwarz-P and Gyroid. Furthermore, we compare retention across different TPMS geometries for both cohesive and cohesionless systems. Finally, for Schwarz-P and Gyroid, we analyse detailed particle kinematics, contact force distribution, and overlap statistics during the discharge.

\section{Simulation Procedure}
In this section, we first provide an overview of the computational framework that has been used for simulating the evacuation of metal powder from TPMS structures. We then provide the details of the contact model that has been employed to calculate forces present during the inter-particle contacts. Next, we establish the cohesive forces and the corresponding particle overlaps produced by varying the parameter in our cohesion model. We finally outline the procedure to create the TPMS  structure embeddings inside a cubic packing, which will act as walls in the the subsequent simulations.

\subsection{The Discrete Element Method}
The Discrete Element Method is a simulation technique in which the dynamics
of a large ensemble of particles is studied by updating the position, velocity, and force of each individual particle with time.
Consider two spherical grains $i$ and $j$ in contact. The
linear and angular momentum balances for the grain $i$ can be written as 
\begin{equation}
\begin{gathered}
m_i \frac{d \mathbf{v}_i}{d t}=m_i \mathbf{F}_i{ }^b+\mathbf{F}_i{ }^c ~\text{,} \\
I_i \frac{d \boldsymbol{\omega}_i}{d t}=\mathbf{T}_i{ }^c ~\text{,}
\end{gathered}
\end{equation}
where $m_i$ is the mass of particle $i, \mathbf{v}_i$ is the linear velocity of its centre-of-mass, $\mathbf{F}_i{ }^b$ is the body force per unit mass acting on $i$, and $\mathbf{F}_i{ }^c$ is the net contact force exerted by all the grains that are in contact with $i$ at time $t, I_i$ is the moment of inertia with respect to an axis through the centre of the sphere, $\boldsymbol{\omega}_i$ is the angular velocity of $i, \mathbf{T}_i{ }^c$ is the net torque on $i$ due to the tangential component of the contact forces between $i$ and all particles that are in contact with it. The contact force and the torque are given by
\begin{equation}
\begin{aligned}
\mathbf{F}_i{ }^c & =\sum_{\substack{j=1 \\
j \neq i}}^N\left(\mathbf{F}_{i j}{ }^n+\mathbf{F}_{i j}{ }^t\right) \text{,}\\
\mathbf{T}_i{ }^c & =\sum_{\substack{j=1 \\
j \neq i}}^N\left(R_i \mathbf{n}\right) \times\left(\mathbf{F}_{i j}{ }^t\right)\text{.}
\end{aligned}
\label{contact_eq}
\end{equation}

In \cref{contact_eq}, $\mathbf{F}_{i j}{ }^n$ and $\mathbf{F}_{i j}{ }^t$ are the normal and tangential components of the contact forces between $i$ and $j$ at time $t, R_i$ is the radius of particle $i$ and $N$ is the number of particles in contact with particle $i$. The unit vector $\mathbf{n}$ along the line joining the centres of $i$ and $j$ is given by

\begin{equation}
\mathbf{n}=\frac{\mathbf{r}_j-\mathbf{r}_i}{\left|\mathbf{r}_j-\mathbf{r}_i\right|}\text{,}
\end{equation}

where $\mathbf{r}_i$ and $\mathbf{r}_j$ are the position vectors of the centres of mass of $i$ and $j$.

\subsection{Contact model details}
The normal and tangential components of the contact forces can be modelled as the sum
of a conservative, elastic, spring force, and a dissipative, viscous, dashpot force. The normal and tangent interaction force between two
particles, $i$ and $j$, are
    \begin{equation}
    \begin{gathered}
    \boldsymbol{F}_{i j}^n=k_n \boldsymbol{\delta}_{n i j}-\gamma_n \boldsymbol{v}_{n i j}~\text{,}\\
    \boldsymbol{F}_{i j}^t=k_i \boldsymbol{\delta}_{t i j}-\gamma_t \boldsymbol{v}_{t i j}~\text{,}
    \end{gathered}    
    \end{equation}
where $\boldsymbol{\delta}_n$ is the overlap of two particles and $\boldsymbol{\delta}_t$ is the tangential displacement of two particles in contact. $\boldsymbol{v}_n$, and $\boldsymbol{v}_t$ are the relative velocity of particles in normal and tangential directions. $k_n$ and $k_t$ are normal and tangential elastic constants.\\
The tangential force is limited by the Coulomb friction criterion,
\begin{equation}
\lVert \boldsymbol{F}_{ij}^t \rVert \leq \mu\, \lVert \boldsymbol{F}_{ij}^n \rVert~\text{,}
\end{equation}
where $\mu$ is the coefficient of friction. When the computed tangential force exceeds this limit, sliding occurs, and the tangential force is truncated to satisfy the equality.
\\\\
The Hertzian contact theory is adopted as
the contact model to calculate the spring and the damping constants,
    \begin{equation}
    \begin{aligned}
    & k_n=\frac{4}{3} E^* \sqrt{R^* \delta_n} ~\text{,}\\
    & k_t=8 G^* \sqrt{R^* \delta_n}~\text{,}
    \end{aligned}    
    \end{equation}
where $E^*$ is the equivalent Young's modulus, $G^*$ is the equivalent shear modulus and $R^*$ is the equivalent radius given by

\begin{equation}
\begin{gathered}
E^*=\frac{E_i E_j}{E_i\left(1-\nu_j^2\right)+E_j\left(1-\nu_i^2\right)}~, \\
\frac{1}{G^*}=\frac{2\left(2-\nu_i\right)\left(1+\nu_i\right)}{E_i}+\frac{2\left(2-\nu_j\right)\left(1+\nu_j\right)}{E_j}~, \\
R^*=\frac{R_i R_j}{R_i+R_j}~.
\end{gathered}
\end{equation}
$\gamma_n$ and $\gamma_t$ are normal and tangential viscoelastic damping constants
\begin{equation}
\begin{aligned}
\gamma_n & =2 \sqrt{\frac{5}{6}} \beta \sqrt{S_n m^*} ~\text{,}\\
\gamma_t & =2 \sqrt{\frac{5}{6}} \beta \sqrt{S_t m^*}~\text{,}
\end{aligned}
\end{equation}
where $S_n, S_t$ and $\beta$ are
\begin{equation}
\begin{aligned}
S_n=2 E^* \sqrt{R^* \delta_n}~\text{,} \\
S_t=8 G^* \sqrt{R^* \delta_n}~\text{,} \\
\beta=\frac{\ln (e)}{\sqrt{\ln ^2(e)+\pi^2}}~\text{.}
\end{aligned}
\end{equation}
$e$ is the coefficient of restitution and $m^*$ is the equivalent mass
\begin{equation}
m^*=\frac{m_i m_j}{m_i+m_j}~\text{.}
\end{equation}
The cohesion between particles is modelled using the SJKR model provided by LIGGGHTS \citep{kloss2012models} as 
\begin{equation}
F=k_c A  ~\text{,} 
\end{equation}
where $k_c$ is the cohesion energy density and $A$ is the contact area of two particles given by 
\begin{equation}
A=\pi~(2R^*\delta_n-\delta_n^2/4)~\text{.}
\end{equation}
        
\begin{figure*}[h]
      \centering           
      \begin{subfigure}{0.49\textwidth}
        \centering
        \input{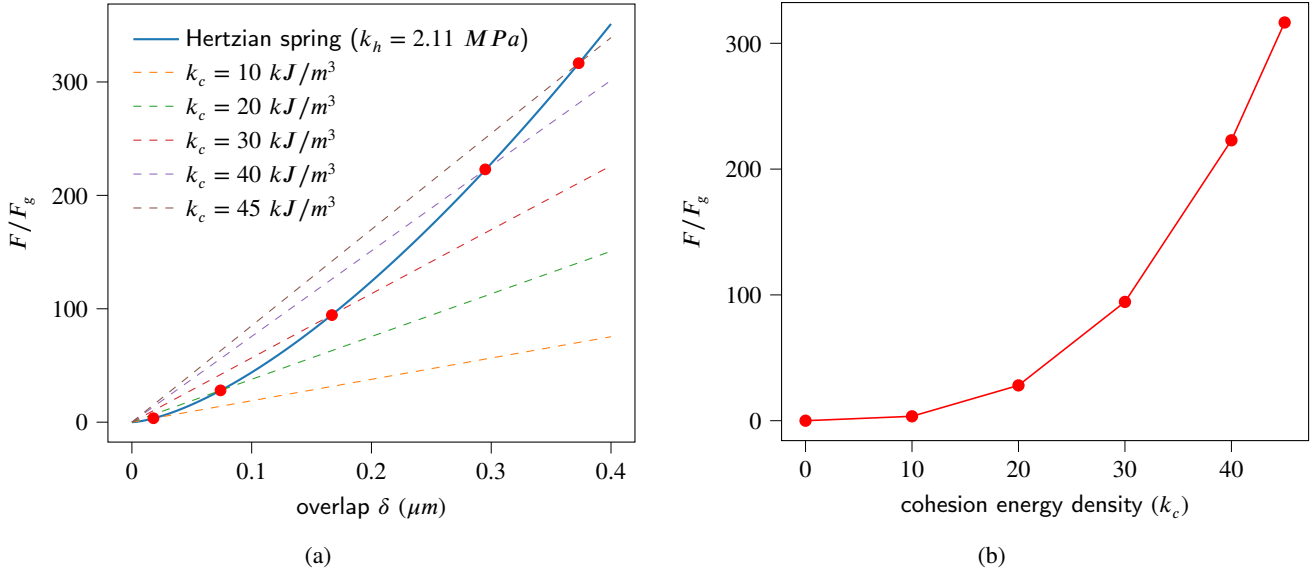}
        \caption{}
      \end{subfigure}           
      \hfill
      \begin{subfigure}{0.49\textwidth}
        \centering
\begin{tikzpicture}

\definecolor{darkgray176}{RGB}{176,176,176}
\definecolor{lightgray204}{RGB}{204,204,204}
\definecolor{steelblue31119180}{RGB}{31,119,180}

\begin{axis}[
legend style={fill opacity=0, draw opacity=1, text opacity=1, draw=none},
tick align=outside,
tick pos=left,
width=\textwidth,
xlabel={cohesion energy density $(k_c)$},
ylabel={$F/F_g$},
x grid style={darkgray176},
xmin=-2.25, xmax=47.25,
xtick style={color=black},
y grid style={darkgray176},
ymin=-15.8278, ymax=332.3838,
ytick style={color=black}
]
\addplot [semithick, red, mark=*, mark size=2, mark options={solid}, forget plot]
table {%
0 0
10 3.484
20 28.056
30 94.439
40 222.908
45 316.556
};
\end{axis}

\end{tikzpicture}
        \caption{}
      \end{subfigure}      
      \caption{Cohesion energy density and the corresponding equilibrium contact force for a given Hertzian spring. (a) shows the equilibrium overlaps and contact forces for two identical spheres modelled as Hertzian springs of stiffness 2.11 $MPa$ at various cohesive energy densities, $k_c$, varying from 0 to 45 $kJ/m^3$, as red dots. (b) shows the variation of the normalised equilibrium forces as a function of $k_c$.}
      \label{cohesion_curve}
    \end{figure*}

        The simulated spheres are modelled as gas-atomised 316 L stainless steel particles. The parameters used in our simulations (presented in \cref{tab:DEM}) are taken
        from \citet{nan2018jamming}, but a cohesion energy density ($k_c$) is specified as opposed to the surface energy parameter used in their paper. A scaled-down Young's modulus is chosen to permit a larger collision time, permitting a larger simulation timestep, $t_s$, and consequently a reduced overall simulation time. For the chosen particle stiffness of $k_h=2.11$ MPa, the Hertzian repulsive force sharply increases with the overlap distance as shown by the solid blue curve in \cref{cohesion_curve}(a). The linear dashed lines show the dependence of the attractive force between a pair of identical particles on their overlap distance due to cohesion. The points of intersection of the solid and  the dashed lines, highlighted as red dots, show the equilibrium overlap distances and the corresponding contact forces. Thus, under static conditions, the maximum overlap is well within 1\% of the particle diameter, even for the most cohesive case employed in this study. The explicit correspondence between $k_c$ (varied systematically in the article) and the ensuing equilibrium contact force is presented in \cref{cohesion_curve}(b). The graph shows how rapidly the forces increase with $k_c$, justifying the chosen limit of 45 $kJ/m^3$ for it in the study. The corresponding Bond number is in accordance with the choices made in \cite{roy2024structural, he2020linking}.
        
            \begin{table*} [h]
                \centering
                \begin{tabular}{|c| c| c|}
                    \hline
                    \textbf{Parameter} & \textbf{Description} & \textbf{Value }\\[0.5 ex]
                    \hline
                    $d_p$ & particle diameter & 45$\times 10^{-6}~m$\\
                    $\rho_p$ & particle density & 7980 $kg/m^3$ \\
                    $E$ & Young's modulus & $2.11~MPa$\\
                    $\nu$ & Poisson ratio & 0.3\\
                    $\mu$ & coefficient of friction & 0.5\\
                    $\mu_r$ & coefficient of rolling friction & 0.0\\
                    $e$ & coefficient of restitution & 0.64 \\
                    $k_c$ & cohesion energy density & \{0, 10, 20, 25, 30, 35, 37, 38, 39, 40, 45\}\\
                    $t_s$ & simulation timestep & $10^{-6}~s$ \\
                    \hline
                \end{tabular}
                \caption{Discrete Element Method (DEM)-specific parameters used in the simulations}
                \label{tab:DEM}
            \end{table*}    

\subsection{Packing and shell generation}           
The common TPMS structures that have been studied from a heat-exchanger application perspective are based on Schwarz-P, Gyroid , and Diamond unit cells. The edge length of a unit cell in these studies \citep{catchpole2019thermal,attarzadeh2021design,qureshi2021heat,attarzadeh2022multi} ranges from 3.3 mm to  10 mm. Similar range of sizes (6 mm to 12 mm) has been employed for energy-absorption lattices\citep{saleh2022compression}. However, the edge length is smaller for tissue engineering applications with values in the range 1.5 mm to 3 mm \citep{wang2025tpms}, with pore sizes designed to mimic mimic cortical or trabecular bones \citep{bioengineering9100504}.  In the present work, the edge length was chosen to be equal to 100 times the particle diameter (45 $\mu m$), making it 4.5 mm.   

\begin{figure*}[h]
      \centering           
      \begin{subfigure}{0.3\textwidth}
        \centering
        \includegraphics[width=\linewidth]{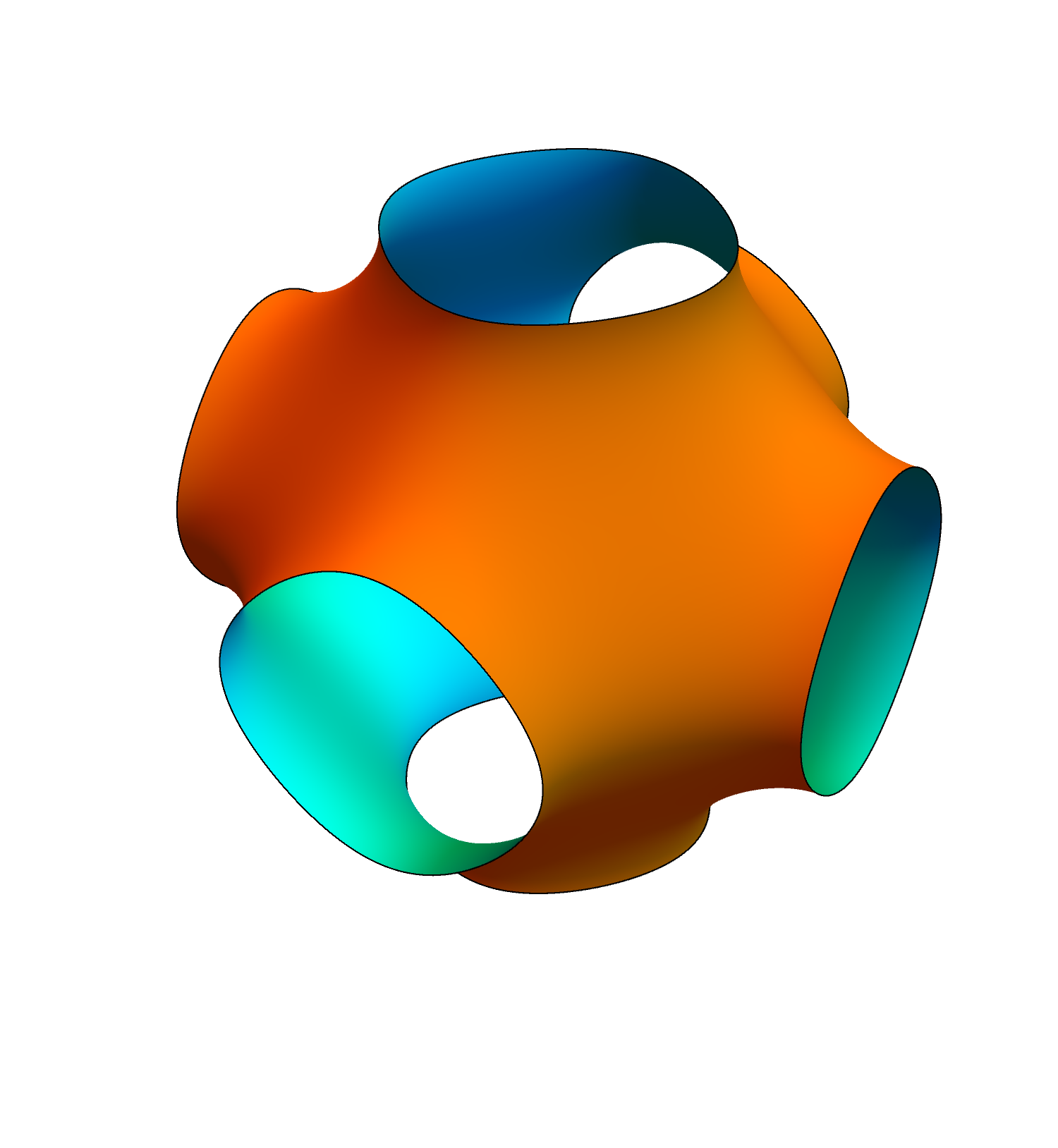}
        \caption{Schwarz-P}
      \end{subfigure}           
      \hfill       
      \begin{subfigure}{0.3\textwidth}
        \centering
        \includegraphics[width=\linewidth]{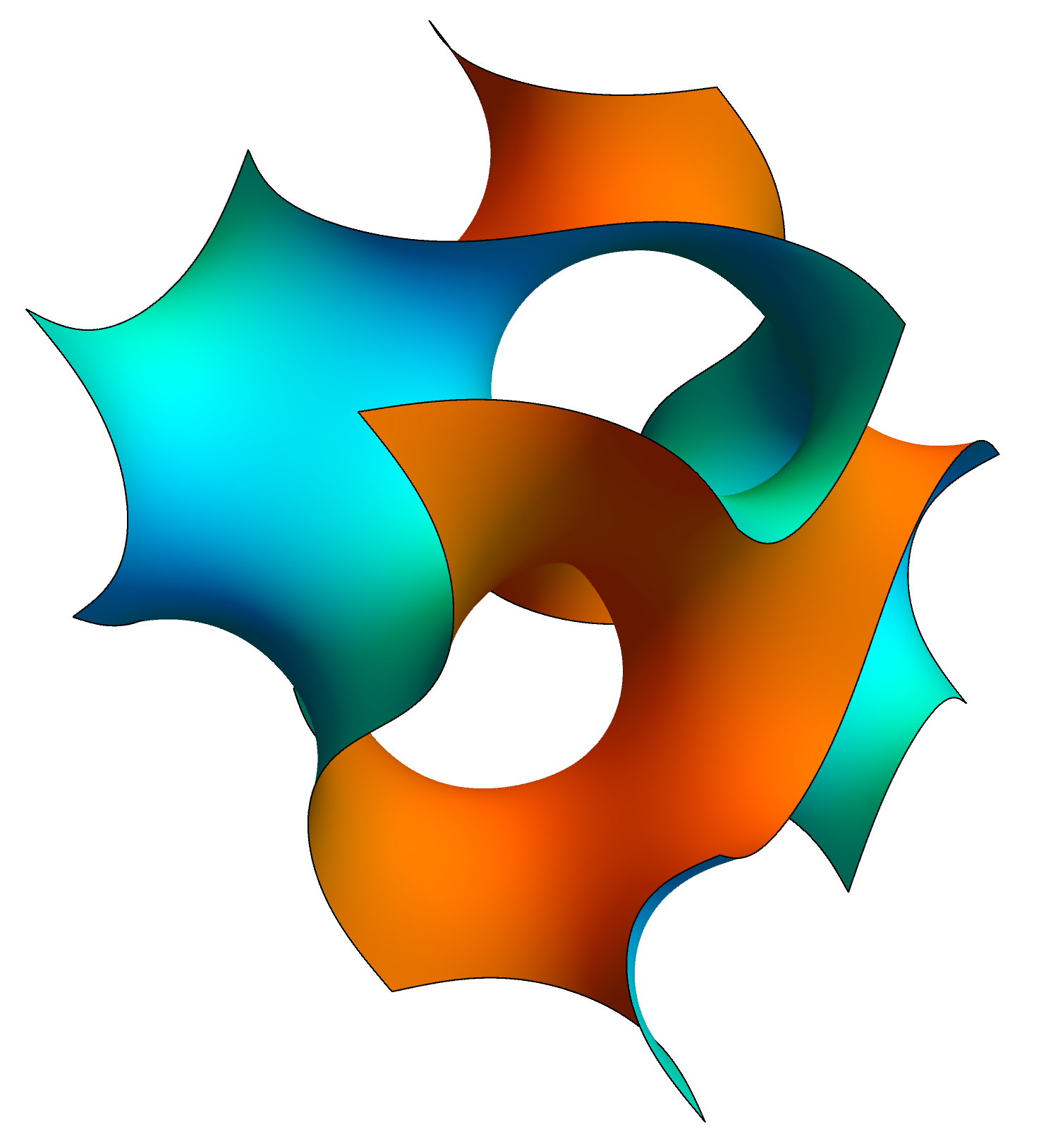}
        \caption{Gyroid}
      \end{subfigure}
      \hfill        
      \begin{subfigure}{0.3\textwidth}
        \centering
        \includegraphics[width=\linewidth]{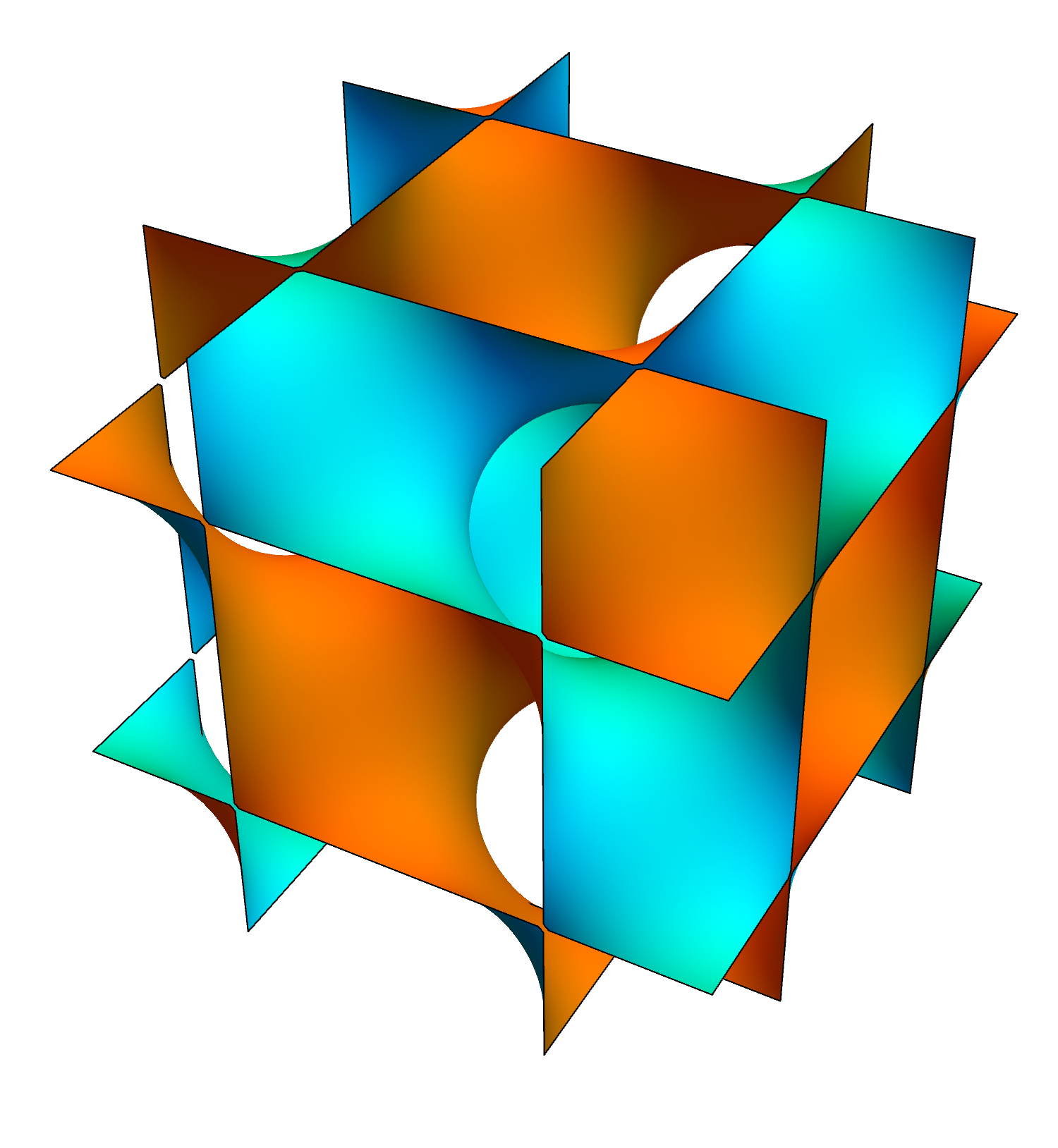}
        \caption{Diamond}
      \end{subfigure}

      \vspace{2em}
      \begin{subfigure}{0.3\textwidth}
        \centering
        \includegraphics[width=\linewidth]{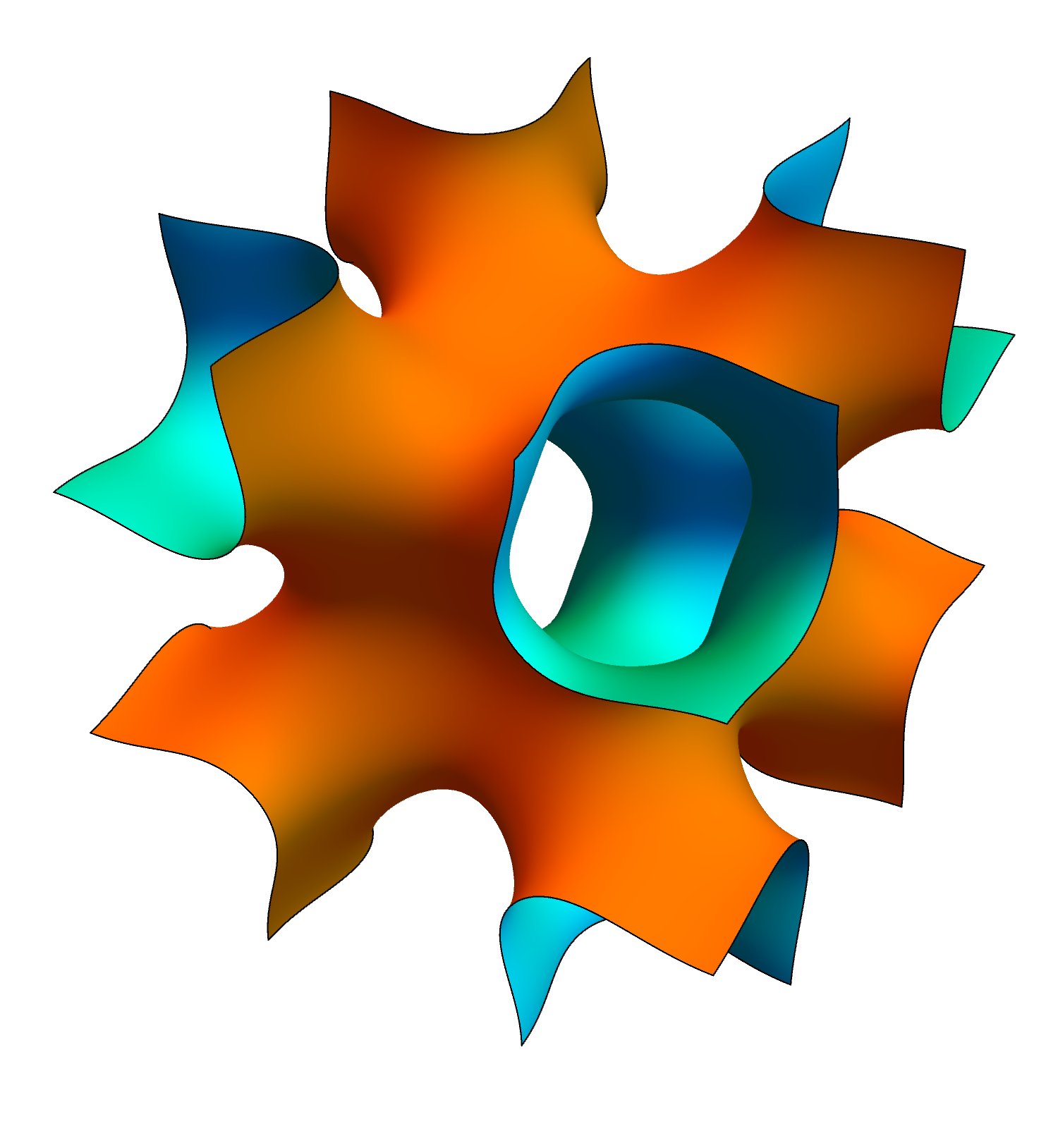}
        \caption{I-WP}
      \end{subfigure}
      \hfill 
      \begin{subfigure}{0.3\textwidth}
        \centering
        \includegraphics[width=\linewidth]{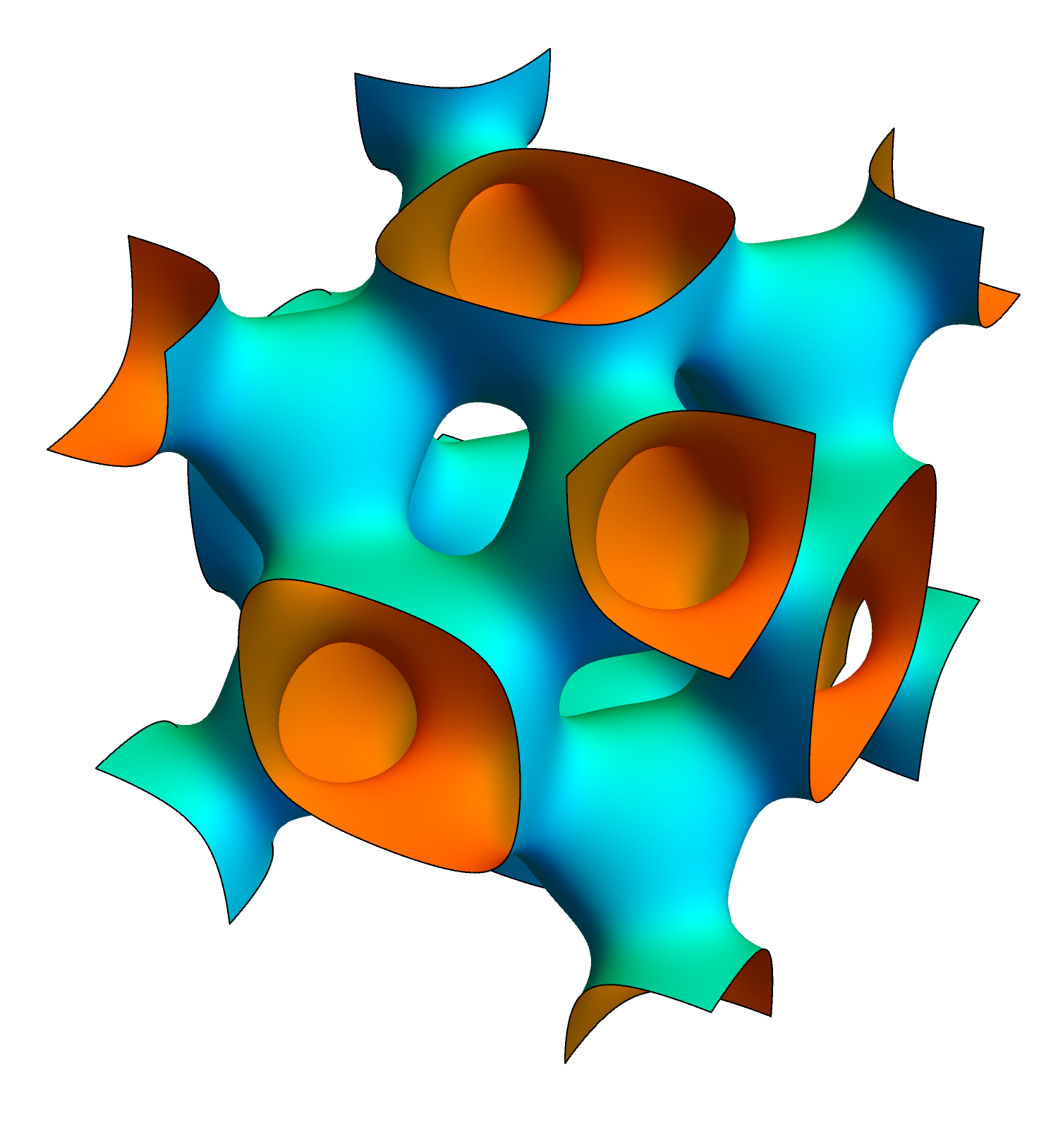}
        \caption{F-RD}
      \end{subfigure}
      \hfill 
      \begin{subfigure}{0.3\textwidth}
        \centering
        \includegraphics[width=\linewidth]{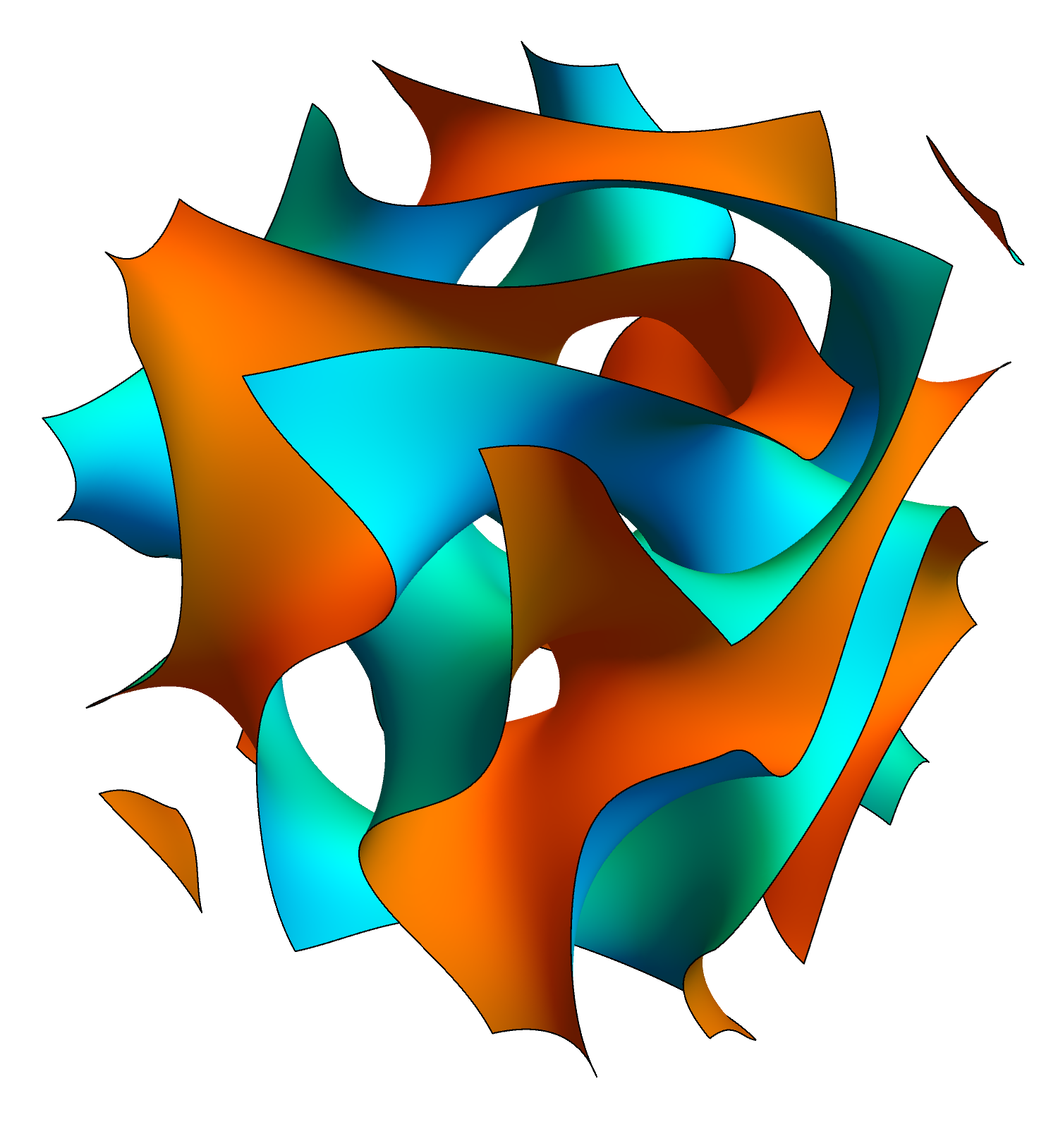}
        \caption{$I_2-Y^{**}$}
      \end{subfigure}
    
      \caption{Various Triply Periodic Minimal Surface (TPMS) unit cells, with their names shown in the respective sub-cations. The colours cyan and orange demarcate the disjoint regions of space partitioned by the surfaces.}
      \label{TPMS_shell}
    \end{figure*}

As a first step to simulate the evacuation of a TPMS unit cell, we need to create a cubic packing where the structure would have been embedded in the case of the actual AM process. In the AM bed-forming stage, a typical layer is only about 2-4 $d_p$ in thickness. Thus, incrementally creating a cubic packing with an edge of 100$d_p$ would need about 30 iterations. We instead directly create the final packing by randomly seeding particles in the cubic region, and then allowing them to relax to the desired packing fraction of 0.62. Next, a subset of particles inside the packing is frozen in space and time to collectively act as a rigid wall, while the other particles can fall freely under gravity. The frozen particles are located exactly where the melting, followed by solidification, would have occurred inside the bulk during the AM process, i.e., the relevant TPMS surface. All TPMS surfaces emerge as solutions to variational problems involving surface area minimization. They are elegantly described by equations (taken from \cite{feng2022triply}), as shown in \cref{TPMS}. The corresponding surfaces themselves are shown in \cref{TPMS_shell}

If $f(\overline{x},\overline{y},\overline{z})=C_1=0$ is the equation of the TPMS surface of interest, two equidistant surfaces $f(\overline{x},\overline{y},\overline{z})=C_2$ and $f(\overline{x},\overline{y},\overline{z})=C_3$ are identified such that the separation between them is only $4 d_p$. $C_2$ can be shown to be $\sin\left(\frac{\pi(4d_p)}{2a}\right)$ and $-\sqrt{2}\sin\left(\frac{\pi(4d_p)}{2a}\right)$ for Schwarz-P and Gyroid respectively, where $a$ is the half-length of the edge of the unit-cell; and $C_3$ is simply $-C_2$. (See Appendix for the derivation.) The particles that lie between these bounding surfaces constitute the TPMS shell to be held stationary during the simulations. The particles on any one side of this shell are deleted for computational efficiency, and only the ones on the other side are retained to be studied in the simulations, nonetheless leaving domain sizes of approximately $5\times 10^5$ particles.

    \begin{table*}[h]
        \centering
        \begin{tabular}{|c| c|}
            \hline
             \textbf{TPMS type} & \textbf{Equation of the surface}  \\[0.5 ex]
             \hline 
             Schwarz-P & $f(\overline{x},\overline{y},\overline{z})=\cos \overline{x}+ \cos \overline{y}+ \cos \overline{z}=C$ \\[1 ex]
             Gyroid & $f(\overline{x},\overline{y},\overline{z})=\sin \overline{x} \cos \overline{y}+ \sin \overline{z} \cos \overline{x}+ \sin \overline{y} \cos \overline{z}=C$\\[1 ex]
             Diamond & $f(\overline{x},\overline{y},\overline{z})=\cos \overline{x} \cos \overline{y} \cos \overline{z}- \sin \overline{x} \sin \overline{y} \sin \overline{z}=C$\\[1 ex]
             I-WP & $f(\overline{x},\overline{y},\overline{z})=2(\cos \overline{x} \cos \overline{y}+ \cos \overline{z} \cos \overline{x}+ \cos \overline{y} \cos \overline{z})-$\\
             &$(\cos 2\overline{x}+ \cos 2\overline{y}+ \cos 2\overline{z})=C$ \\[1 ex]
             F-RD & $f(\overline{x},\overline{y},\overline{z})=4\cos \overline{x} \cos \overline{y} \cos \overline{z}-$\\
             &$(\cos 2\overline{x} \cos 2\overline{y}+ \cos 2\overline{z} \cos 2\overline{x}+ \cos 2\overline{y} \cos 2\overline{z})=C$ \\[1 ex]
             $I_2-Y^{**}$ & $f(\overline{x},\overline{y},\overline{z})=-2(\sin 2\overline{x} \cos \overline{y} \sin \overline{z}+ \sin 2\overline{z} \cos \overline{x} \sin \overline{y}+ \sin 2\overline{y} \cos \overline{z} \sin \overline{x})+$ \\
             & $(\cos 2\overline{x} \cos 2\overline{y}+ \cos 2\overline{z} \cos 2\overline{x}+ \cos 2\overline{y} \cos 2\overline{z})=C$\\
             \hline
        \end{tabular}
        \caption{TPMS cell types and their equations of surface represented as $f(\overline{x},\overline{y},\overline{z})=C$, where $\overline{x}=\pi (x-a)/a$, $\overline{y}=\pi (y-a)/a$ , and $\overline{z}=\pi (z-a)/a$, $a$ being half-length of the edge of the unit-cell. $C$ was taken as zero in our study, as it divides the cube into regions of equal volume.}
        \label{TPMS}
    \end{table*}
    
\section{Results and Discussion}
\begin{figure*}[h]
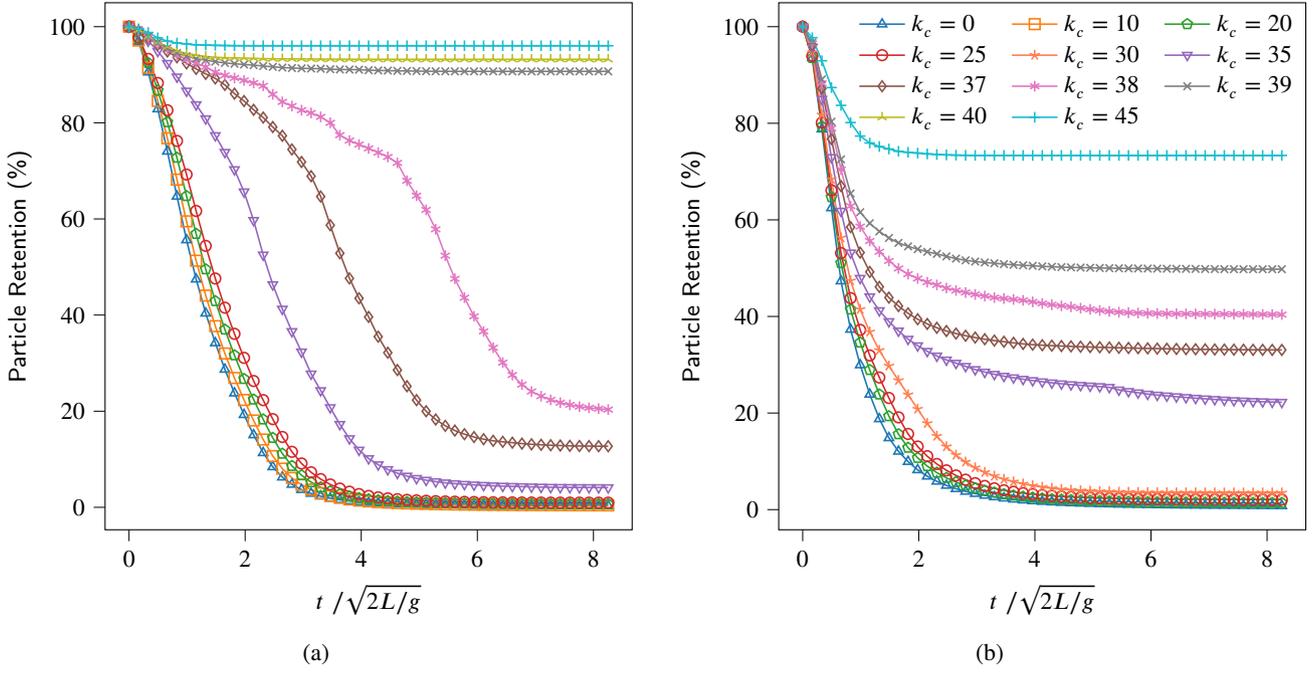

  \centering           
  \begin{subfigure}{0.49\textwidth}
    \centering
    \input{Figures/misc/sch_p}
    \caption{}
  \end{subfigure}           
    \hfill
  \begin{subfigure}{0.49\textwidth}
    \centering
    \input{Figures/misc/gyroid}
    \caption{}
  \end{subfigure}           
  \caption{Variation of particle retention inside the unit cell with normalised time for (a) Schwarz-P and (b) Gyroid structures at different cohesive energy densities. $k_c$ varies from 0 to 45 $kJ/m^3$. The common legend is shown in (b).}
  \label{retention_cohesion}
\end{figure*}

\begin{figure*}[H]
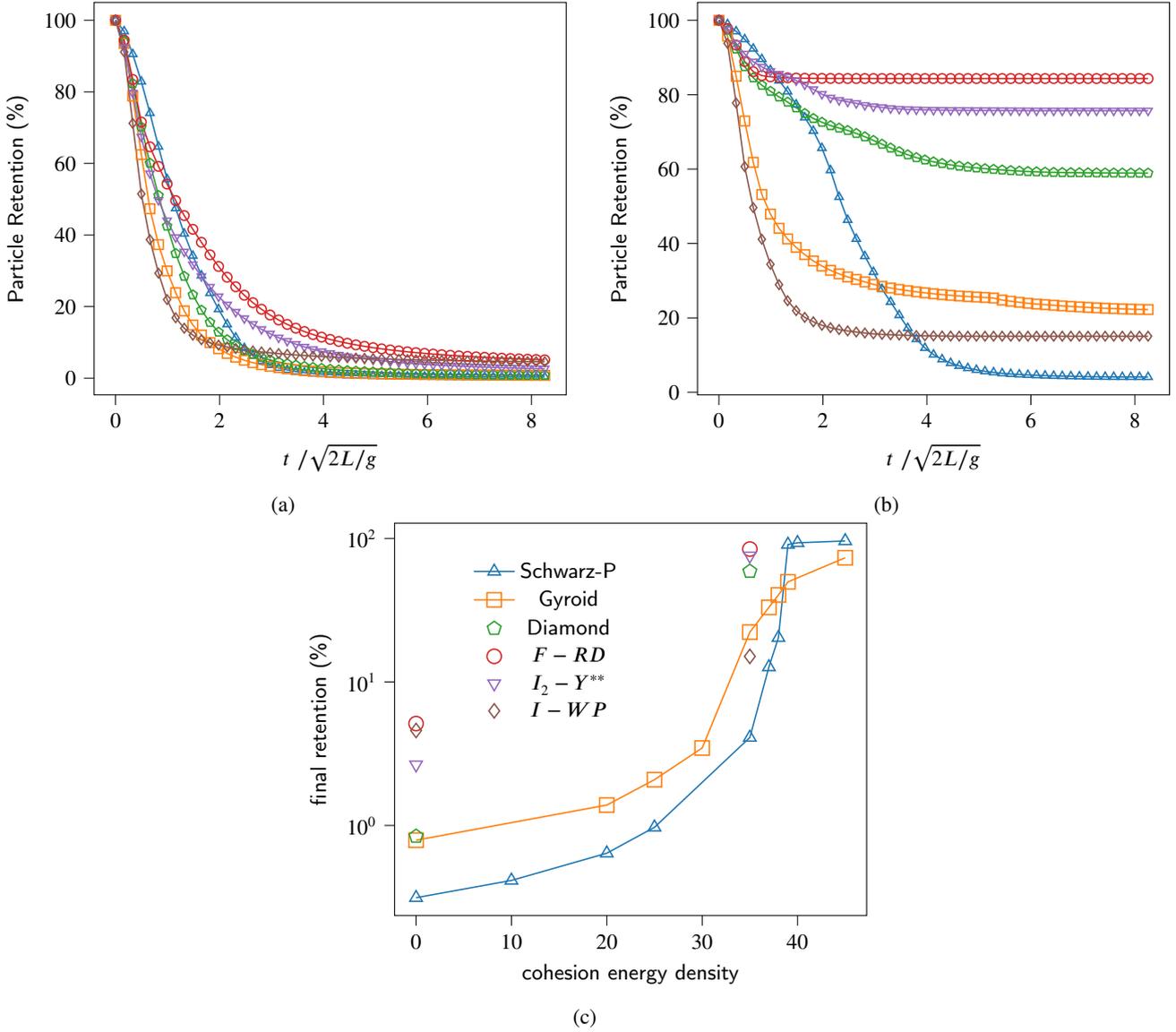

  \centering           
  \begin{subfigure}{0.49\textwidth}
    \centering
    \input{Figures/misc/all_TPMS.tex}
    \caption{}
  \end{subfigure}           
    \hfill
  \begin{subfigure}{0.49\textwidth}
    \centering
    \input{Figures/misc/all_TPMS_coh.tex}
    \caption{}
  \end{subfigure}           
  \\
  \begin{subfigure}{0.49\textwidth}
    \centering
\begin{tikzpicture}

\definecolor{crimson2143940}{RGB}{214,39,40}
\definecolor{darkgray176}{RGB}{176,176,176}
\definecolor{darkorange25512714}{RGB}{255,127,14}
\definecolor{forestgreen4416044}{RGB}{44,160,44}
\definecolor{lightgray204}{RGB}{204,204,204}
\definecolor{mediumpurple148103189}{RGB}{148,103,189}
\definecolor{sienna1408675}{RGB}{140,86,75}
\definecolor{steelblue31119180}{RGB}{31,119,180}

\begin{axis}[
legend style={fill opacity=0, draw opacity=1, text opacity=1, at={(0.15,0.7)}, anchor=west, draw=none},
log basis y={10},
tick align=outside,
tick pos=left,
width=\textwidth,
x grid style={darkgray176},
xlabel={cohesion energy density},
xmin=-2.25, xmax=47.25,
xtick style={color=black},
y grid style={darkgray176},
ylabel={final retention (\%)},
ymin=0.235719055398696, ymax=127.85558911115,
ymode=log,
ytick style={color=black},
ytick={0.01,0.1,1,10,100,1000,10000},
yticklabels={
  \(\displaystyle {10^{-2}}\),
  \(\displaystyle {10^{-1}}\),
  \(\displaystyle {10^{0}}\),
  \(\displaystyle {10^{1}}\),
  \(\displaystyle {10^{2}}\),
  \(\displaystyle {10^{3}}\),
  \(\displaystyle {10^{4}}\)
}
]
\addplot [semithick, steelblue31119180, mark=triangle, mark size=3, mark options={solid,fill opacity=0}]
table {%
0 0.313821138211382
10 0.414092140921409
20 0.640831074977416
25 0.971093044263776
35 4.10135501355014
37 12.7080397470641
38 20.3380307136405
39 90.7031616982837
40 93.2099367660343
45 96.0355916892502
};
\addlegendentry{Schwarz-P}

\addplot [semithick, darkorange25512714, mark=square, mark size=3, mark options={solid,fill opacity=0}]
table {%
0 0.789611772604156
20 1.3868846892869
25 2.08428184490968
30 3.46460733549916
35 22.2354906666461
37 33.0680073154357
38 40.3892305674093
39 49.7856685366814
45 73.3434164936839
};
\addlegendentry{Gyroid}

\addplot [semithick, forestgreen4416044, mark=pentagon, mark size=3, mark options={solid,fill opacity=0}, only marks]
table {%
0 0.84
35 58.93
};
\addlegendentry{Diamond}

\addplot [semithick, crimson2143940, mark=o, mark size=3, mark options={solid,fill opacity=0}, only marks]
table {%
0 5.13
35 84.33
};
\addlegendentry{$F-RD$}

\addplot [semithick, mediumpurple148103189, mark=triangle, mark size=3, mark options={solid,rotate=180,fill opacity=0}, only marks]
table {%
0 2.65
35 75.69
};
\addlegendentry{$I_2-Y^{**}$}

\addplot [semithick, sienna1408675, mark=diamond, mark size=3, mark options={solid,fill opacity=0}, only marks]
table {%
0 4.59
35 15.07
};
\addlegendentry{$I-WP$}

\end{axis}

\end{tikzpicture}
    \caption{}
  \end{subfigure}           
  \caption{Effect of geometry on particle retention. Shown in
  (a) is the retention profile with normalised time for cohesionless particles,  and in (b) the profile for a cohesive energy density $(k_c)~=35~ kJ/m^3$. (c) shows the final retention in the different geometries as a function of $k_c$. The common legend for all the sub-panels is present in (c).}
  \label{retention_geometry}
\end{figure*}

\begin{figure*}[H]
\centering
\includegraphics[width=\linewidth]{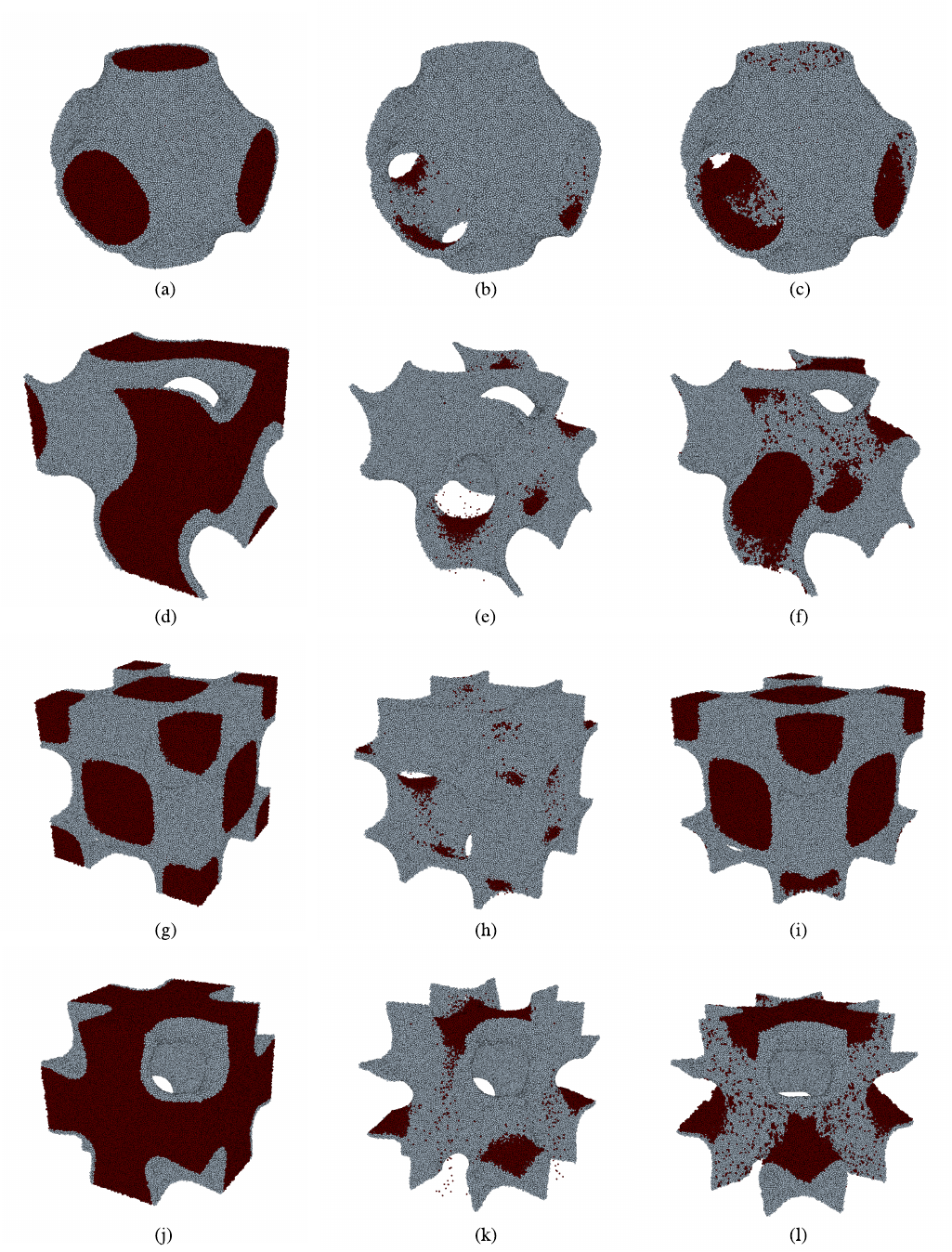}
      \caption{Initial and final retention for cohesionless and cohesive powder for Schwarz-P, Gyroid, Diamond, and I-WP unit cells in rows 1,2,3 and 4 respectively. The first column shows the configuration at the start of the simulation. The gray particles constitute the shell and the maroon particles are the trapped ones. The second column shows the trapped particles at the end of the evacuation for $k_c~=0~ kJ/m^3$ , and the third column for $k_c~=35~ kJ/m^3$.}
      \label{all_retention_pic}
\end{figure*}

\begin{figure*}[H]
\centering
\includegraphics[width=\linewidth]{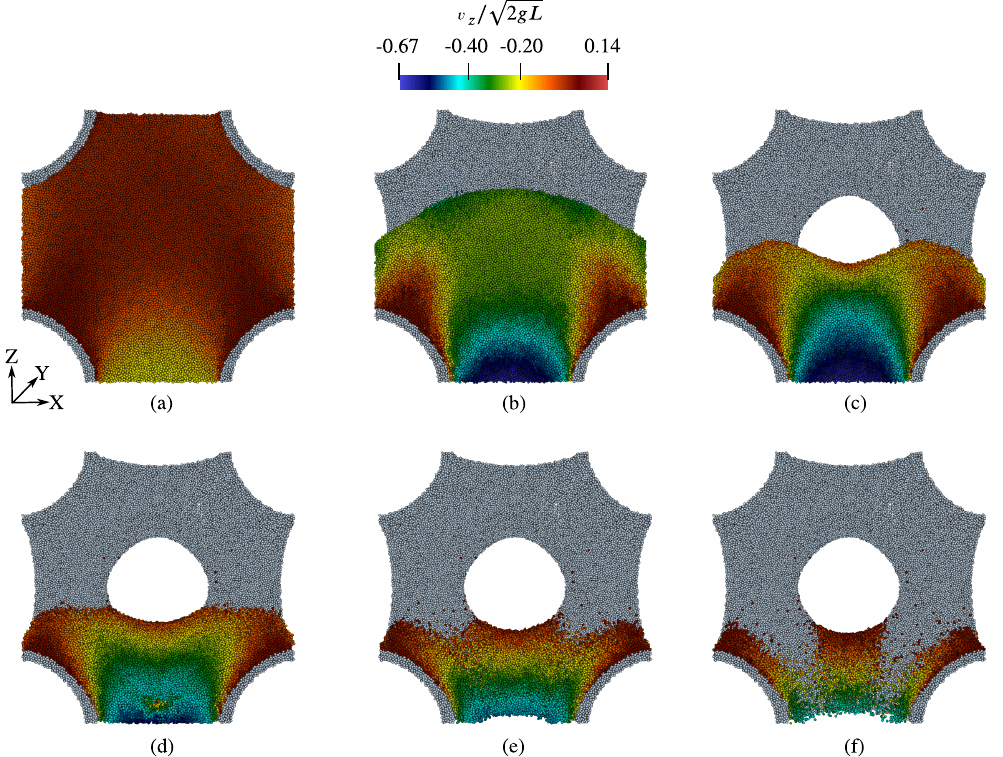}
      \caption{Stages in the natural evacuation of a Schwarz-P unit cell under gravity (acting along negative $z$-direction). (a), (b), (c), (d), (e) and (f) correspond to the elapsed times of $t/\sqrt{2L/g}= 0.165,~0.825,~1.49,~2.15,~2.81,~3.47$ from the start of evacuation, respectively. The free particles are coloured based on the normalised z-component of their velocities $v_z/\sqrt{2gL}$, with the wall particles being represented in gray. }
      \label{sch_p_vel}
\end{figure*}

\begin{figure*}[H]
\centering
\includegraphics[width=\linewidth]{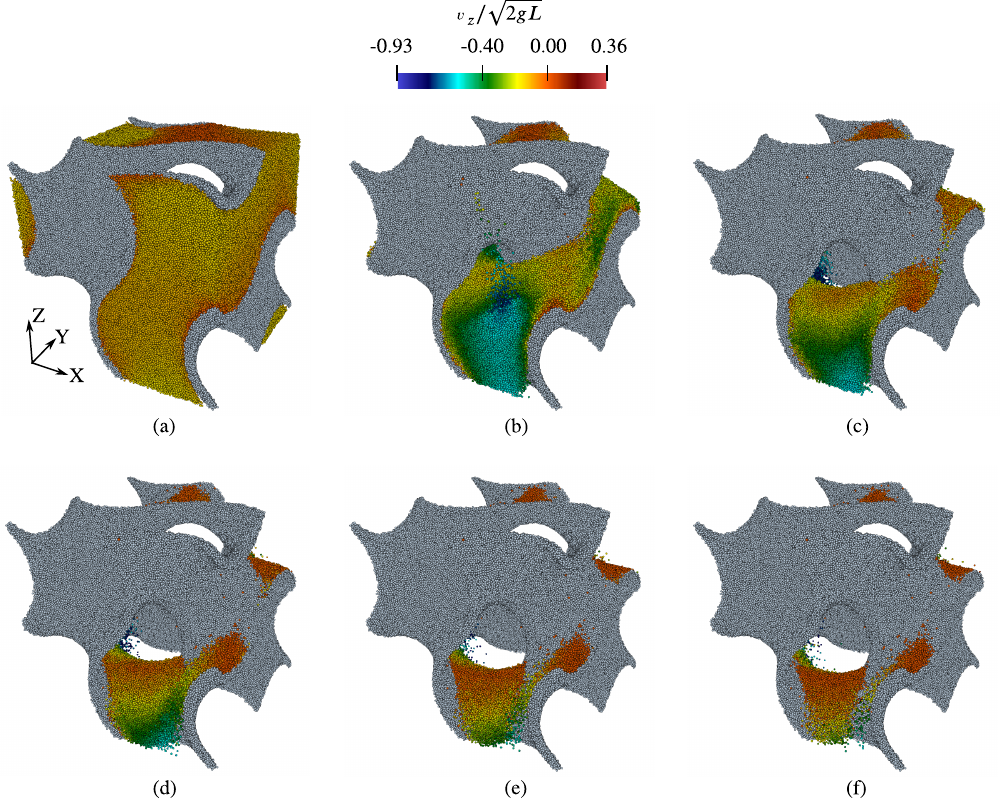}
      \caption{Stages in the natural evacuation of a Gyroid unit cell under gravity (acting along negative $z$-direction). (a), (b), (c), (d), (e), and (f) correspond to the elapsed times of $t/\sqrt{2L/g}= 0.165,~0.825,~1.49,~2.15,~2.81,~3.47$ from the start of evacuation, respectively. The free particles are coloured based on the normalised z-component of their velocities $v_z/\sqrt{2gL}$, with the wall particles being represented in gray.}
      \label{gyroid_vel}
\end{figure*}

\begin{figure*}[H]
  \centering           
  \begin{subfigure}{0.49\textwidth}
    \centering
    \input{Figures/misc/histogram/sch_p_delta_0_1}
    \caption{}
  \end{subfigure}   
  \hfill    
  \begin{subfigure}{0.49\textwidth}
    \centering
    \input{Figures/misc/histogram/sch_p_force_0}
    \caption{}
  \end{subfigure}           
  \\
  \begin{subfigure}{0.49\textwidth}
    \centering
    \input{Figures/misc/histogram/sch_p_delta_35}
    \caption{}
  \end{subfigure}           
    \hfill 
  \begin{subfigure}{0.49\textwidth}
    \centering
    \input{Figures/misc/histogram/sch_p_force_35}
    \caption{}
  \end{subfigure}           
    \caption{Histogram of the normalised deformation and the normalised contact force for Schwarz-P TPMS cell at $k_c~=0~ kJ/m^3$ [(a) and (b)] and at $k_c~=35~ kJ/m^3$ [(c) and (d)]. The different colours correspond to different instants of time as shown in the legend at the top.}
  
   \label{histo_sch_p}
    \end{figure*}

\begin{figure*}[H]
  \centering           
  \begin{subfigure}{0.49\textwidth}
    \centering
    \input{Figures/misc/histogram/gyroid_delta_0_1}
    \caption{}
  \end{subfigure}   
  \hfill    
  \begin{subfigure}{0.49\textwidth}
    \centering
    \input{Figures/misc/histogram/gyroid_force_0}
    \caption{}
  \end{subfigure}           
  \\
  \begin{subfigure}{0.49\textwidth}
    \centering
    \input{Figures/misc/histogram/gyroid_delta_35}
    \caption{}
  \end{subfigure}           
    \hfill 
  \begin{subfigure}{0.49\textwidth}
    \centering
    \input{Figures/misc/histogram/gyroid_force_35}
    \caption{}
  \end{subfigure}           
    \caption{Histogram of the normalised deformation and the normalised contact force for Gyroid TPMS cell at $k_c~=0~ kJ/m^3$ [(a) and (b)] and at $k_c~=35~ kJ/m^3$ [(c) and (d)]. The different colours correspond to different instants of time as shown in the legend at the top.}
  
   \label{histo_gyroid}
    \end{figure*}

We next examine how varying the cohesive strength influences the discharge profiles in the TPMS unit cells. For Schwarz-P, when the cohesive strength is turned up from 0 to 25 $kJ/m^3$, as shown in \cref{retention_cohesion}(a), there is minimal change in the discharge profile. These cases are characterized by a steep evacuation from the beginning, continuing until the 10\% retention point occurring at $t/\sqrt{2L/g}=2.5$, around which it drastically slows down. As $k_c$ is increased to $35~kJ/m^3$, the early discharge rate slows down, resuming the steep profile only after 30\% evacuation. The onset of the rapid discharge is delayed from $t/\sqrt{2L/g}=2$ to 3 to 5, as $k_c$ changes from 35 to 37 to 38 $kJ/m^3$. There is also a noticeable rise in the final steady-state retention percentage during the change in $k_c$. For $k_c=$ 39, 40 and $45~kJ/m^3$, a slow early brief discharge is followed by stagnation retaining more than 90\% of particles inside the cell.

The influence of $k_c$ on the discharge profile is qualitatively different in the Gyoid unit cell as compared to the Schwarz-P unit cell, as seen in \cref{retention_cohesion}(b). This is because the early discharge rate is not a function of cohesion for the Gyroid. Furthermore, much of the dynamics at all cohesive strengths occurs before $t/\sqrt{2L/g}=2$, at which point the retention is very close to the eventual steady-state values. Similar to Schwar-P, low values of cohesion, $k_c\leq 30~kJ/m^3$, all result in almost identical discharge profiles. The final retention rate takes a significant jump from $k_c=$ 30 to 35 $kJ/m^3$, and continues to be responsive to further increments in $k_c$ until the simulated value of 45$kJ/m^3$.  

Comparing the discharge profiles of cohesionless particles across various TPMS unit cells as shown in \cref{retention_geometry}, we observe some interesting differences. While the Schwarz-P, the Gyroid, and the diamond cells settle to almost similar final retention percentages viz. 0.31\%, 0.79\%, and 0.84\% respectively, their initial discharge dynamics vary noticeably as shown in \cref{retention_geometry}(a). The F-RD and I-2Y start off similar to the previous three structures but quickly slow down to lower discharge rates, finally retaining 5.13\% and 2.65\% of material, respectively. The I-WP stands out as it has the fastest discharge rate amongst all up until the 90\% evacuation point, beyond which it dramatically slows down, retaining 4.59\%, second to F-RD.

The trends are considerably different when it comes to the higher cohesion strength of 35 $kJ/m^3$ as depicted in \cref{retention_geometry}(b). The individual differences among the structures start to become more significant. The I-WP still has the fastest early discharge rate, followed by Gyroid with their eventual retention rates as 15.1\% and 22.2\%, respectively. F-RD, I-2Y and the diamond, are the worst-performing structures retaining more than half of the initially trapped powder. The Schwarz-P, displaying a slow early discharge picking up around $t/\sqrt{2L/g}=2$ and  considerably slowing beyond $t/\sqrt{2L/g}=2$, is the best performing structure retaining merely 4.10\%.

The locations where the powders tend to get trapped can be seen in \cref{all_retention_pic} for the two cohesive strengths. These are usually the sites where the local topology can facilitate via microscopic and geometric friction to counter the effects of particle inertia and the pull due to gravity.

Comparing the final retention percentages for the various TPMS cells across a wide range of $k_c$ (refer \cref{retention_geometry}(c)), we observe Schwarz-P and Gyroid consistently outperform the other structures with respect to ease of recovery of the trapped powder,  The diamond, and the I-WP's performances rely heavily on the cohesion value, making it non-trivial to label one over the other as a better choice.

Next, we closely investigate the kinematics and force transmission in the two consistently favourable unit cells. In \cref{sch_p_vel}, six equispaced snapshots in time reveal the evolution of the z-component of velocity of particles on the $y=0$ plane in a Schwarz-P unit cell. The cell evacuates under the influence of gravity acting along the negative $z$-direction. The velocities have been scaled with the maximum velocity that an isolated freely falling particle would attain while traversing the full edge-length of the cell. This allows us to quantify the overall resistance encountered by the evacuating particles as a combined effect of the inter-particle bulk dynamics and the geometry interference. The velocity field, in the earlier stages, reveals dome-like iso-contours at the exit of the cell, while the upper front transitions from a creeping flat shape in \cref{sch_p_vel}(a) to a rapid convex shape in \cref{sch_p_vel}(b) to a slower biconvex shape with a concave inflexion in \cref{sch_p_vel}(c). This is reminiscent of transient arching discussed in the literature of flow through hoppers and silos. Correspondingly, the maximum normalised peak velocity magnitude first increases from 0.23 in \cref{sch_p_vel}(a) to 0.64 in \cref{sch_p_vel}(b), plateauing at 0.65 until \cref{sch_p_vel}(c), after which it gradually wears off to 0.62 in \cref{sch_p_vel}(d). Beyond \cref{sch_p_vel}(d) the arching structures rapidly collapse marking the transition from a funnel flow to a mass-flow regime rendering a peak velocity magnitude of 0.52 in \cref{sch_p_vel}(e) and of 0.47 eventually in \cref{sch_p_vel}(f). The maximum magnitude of 0.67 achieved during the process would correspond to a free-fall height of $0.45L$, which is roughly the fill height in \cref{sch_p_vel}(c). Throughout the evacuation, particles closer to the walls remain slow-moving or stagnant as a consequence of the frictional resistance from them.

The snapshots of the development of the flow in the Gyroid unit cell shown in \cref{gyroid_vel} contrast those in the Schwarz-P unit cell. The most apparent distinction is that in the scaled maximum velocity magnitude, which attains a value of 0.93 (38.8\% higher than that of the Schwarz-P cell), indicating a much lower resistance offered by the geometry. The value would correspond to a free-fall height of $0.86L$. This can be attributed to the absence of the cellular symmetry that had aided in the formation of arches for Schwarz-P. The maximum velocity attained climbs rapidly from 0.20 in \cref{gyroid_vel}(a) to 0.78 in \cref{gyroid_vel}(b) to 0.87 in \cref{gyroid_vel}(c) at which point a significant fraction of particles has already been expelled. The peak velocity continues to rise reaching 0.91 in \cref{gyroid_vel}(d) to 0.92 in \cref{gyroid_vel}(e), after which it drops slightly to 0.89 owing to the discharge of the particles adjacent to the walls, experiencing an enhanced resistance.

The histograms of the overlaps and the contact forces for cohesionless particles in Schwarz-P follow an exponential-like distribution at all times as shown in \cref{histo_sch_p}(a) and \cref{histo_sch_p}(b). This is also true for the Gyroid structure (see \cref{histo_gyroid}(a) and \cref{histo_gyroid}(b)). The maximal overlap is about 1.4$\%$ of the particle diameter, and the corresponding contact force is $\approx700~F_g$, where $F_g$ is the weight of a single particle. The upper limit of the overlaps and the contact forces are intimately linked to the acceleration of the particles. This is because a cohesionless granular column of $100 d_p$, equal to the height of the unit cell cannot transmit a load greater than $100F_g$, The presence of higher values in fact result from the momentum transferred among the particles due to collisions. Nevertheless, the majority of particles still carry loads below the static limit. Furthermore, the maximum overlaps and contact forces for Schwarz-P increase drastically from $t/\sqrt{2L/g}=$ 0.165 to 0.825, followed by a sharp decline at 1.49. This validates our hypothesis of the resisting arching structures formed to explain the transient slowing down in the evacuation. This is contrary to the monotonic rise observed in the maximum overlaps/contact forces for the Gyroid structure. The contact force histograms for cohesive particles ($k_c=35~ kJ/m^3$) are still exponential-type for both Schwarz-P and Gyroid (comparing \cref{histo_sch_p}(d) and \cref{histo_gyroid}(d)), but cohesion now causes the equilibrium value of the overlaps to be higher than before, as seen in \cref{histo_sch_p}(c) and \cref{histo_gyroid}(c), $\delta/d$ peaking around 0.5\%. Even with such large cohesion, the overlap never exceeds 3\% of the particle diameter, which is true only for a small minority of particles, justifying the choice of the spring stiffness value in the present study.  Cohesion also produces the interesting effect of an earlier arching onset which even sustains for a longer period compared to the cohesionless case.

\section{Conclusion}
This study examined the evacuation of metal powders from TPMS unit cells, focusing on the interplay between geometry and cohesion. We developed a computational framework to pack particles inside TPMS geometries, and analysed the resulting discharge dynamics under gravity. While discharge in silos has been extensively studied, we found that complex pore geometries exhibit distinct kinematic features. In particular, Schwarz-P shows a drastic mid-way slowdown, indicative of transient arching, while Gyroid displays consistently higher particle velocities and less resistance to flow. Force-chain statistics and contact-force distributions further corroborate the presence of arch-like structures and their role in modulating discharge behaviour.

Across geometries, Schwarz-P and Gyroid consistently outperformed other TPMS structures with respect to powder evacuation efficiency. For cohesionless particles, they retained less than 1\% of the trapped powder, while maintaining relatively favourable flow dynamics even at high cohesion levels. In contrast, I-WP, F-RD, and Diamond exhibited significant sensitivity to cohesion, often leading to large residual powder fractions. These results suggest that Schwarz-P and Gyroid are robust design choices for AM applications where de-powdering is critical.

\citet{meier2019modeling} performed DEM simulations of cohesive metal powders in AM and reported that increasing (decreasing) the mean powder particle size by a certain factor has an equivalent effect on the bulk powder behaviour as decreasing (increasing) the surface energy by the square of that factor. For our system, this implies that using finer powders—while desirable for improved surface finish—would effectively amplify cohesive effects, leading to an earlier onset of poor discharge and higher powder retention in TPMS geometries.

Several more complex physical aspects present exciting opportunities for future exploration beyond the scope of this work. First, real AM powders often deviate from perfect sphericity, and \citet{mehrabi2023investigation} demonstrated that shape irregularities can induce ratholing and unpredictable clogging. Second, electrostatic charging—arising from powder–powder or powder–wall interactions—can persist due to insulating oxide layers on metallic particles \cite{alchikh2020powder}. Such effects lead to a complex interplay of attractive and repulsive interactions, making discharge behaviour highly non-trivial. These considerations highlight that our conclusions are specific to idealized spherical, van der Waals–dominated powders.

A natural progression of this study is to investigate de-powdering in periodic TPMS arrays, where long-range structural correlations may further influence flow. An alternative packing generation protocol involving Lees-Edwards shear \citep{lees1972computer,berry2021lees}---arguably a better representation of the bed-forming process---didn't seem to have any impact on the discharge profile of cohesionless powder through the Schwarz-P cell. However, further investigation is required to ascertain this invariance across geometries, and microscopic powder properties. Furthermore, beyond static discharge, external excitations such as vibration or rotation could serve as unjamming strategies, and their systematic evaluation may reveal critical thresholds of frequency and amplitude. Coupled CFD–DEM approaches could also be leveraged to test the efficacy of localized air jets in breaking bridges and accelerating evacuation, inspired by prior studies of air-driven pattern formation in granular beds \cite{maiti2015evolution}. Together, these directions promise to deepen our understanding of powder evacuation in AM and inform geometry-aware design rules for complex porous scaffolds.

\printcredits

\section*{Acknowledgement}
A.K.G. thanks Anushanth Karalasingam and Kevin Hanley for helpful discussions, and the University of Edinburgh for supporting this work through a doctoral studentship.

\subsection*{Data Availability}
All data associated with this work are available at \url{https://github.com/aashish-k-gupta/TPMS-data}.

\bibliographystyle{model1-num-names}
\bibliography{references}

\section*{Appendix}

\subsection*{A. Derivation of the Schwarz--P shell parameters}
\noindent We start with the equation of the Schwarz-p surface
\[
f(\bar{x},\bar{y},\bar{z}) = \cos \bar{x} + \cos \bar{y} + \cos \bar{z} = C, \tag{A1}
\]
where
\[
\bar{x} = \frac{\pi (x - a)}{a}, \quad 
\bar{y} = \frac{\pi (y - a)}{a}, \quad 
\bar{z} = \frac{\pi (z - a)}{a}, 
\]
and
\[
(x, y, z) \in [-a, a]^3~.
\]
\\
Let us first find \( z \) when \( x = a, \, y = 0 \) by substituting the same in Eq. A1.
\begin{align*}
&\cos(0) + \cos(-\pi) + \cos\!\left( \frac{\pi (z - a)}{a} \right) = C \\[4pt]
\Rightarrow \quad &1 - 1 + \cos\!\left( \frac{\pi (z - a)}{a} \right) = C \\[4pt]
\therefore \quad &z = \frac{a}{\pi} \cos^{-1}(C) + a \tag{A2}
\end{align*}
\\
If $f(\bar{x},\bar{y},\bar{z}) = C_2$ and  $f(\bar{x},\bar{y},\bar{z}) = C_3$ are surfaces situated symmetrically about $f(\bar{x},\bar{y},\bar{z}) = C_1=0$ such that the separation between them is $t$, then
\[
\frac{t}{2} = z(C_2) - z(C_1).
\]
Substituting from (2), we get
\begin{align*}
\frac{t}{2} &= \frac{a}{\pi} \left[ \cos^{-1}(C_2) - \cos^{-1}(C_1) \right] \\
\Rightarrow \quad\frac{t}{2} &= \frac{a}{\pi} \left[ \cos^{-1}(C_2) - \cos^{-1}(0) \right]\\
\Rightarrow \quad \frac{t}{2} &= \frac{a}{\pi} \left[ \cos^{-1}(C_2) + \frac{\pi}{2} \right]~.
\end{align*}
If the shell thickness is n-particles wide, then
\begin{align*}
 &\frac{nd_p}{2} = \frac{a}{\pi} \left[ \cos^{-1}(C_2) + \frac{\pi}{2} \right]\\
\Rightarrow \quad&\cos^{-1}(C_2) = \frac{n \pi}{2} \left( \frac{d_p}{a} \right) - \frac{\pi}{2} \\[4pt]
\therefore \quad &C_2 = \cos\!\left[ \frac{n \pi}{2} \left( \frac{d_p}{a} \right) - \frac{\pi}{2} \right] = \sin\!\left( \frac{n \pi}{2} \frac{d_p}{a} \right). \tag{A3}
\end{align*}
\\
Similarly,
\begin{align*}
&\frac{t}{2} = z(C_1) - z(C_3)\\
\therefore \quad &C_3 = -\sin\!\left( \frac{n \pi}{2} \frac{d_p}{a} \right) \tag{A4}
\end{align*}

\subsection*{B. Derivation of the Gyroid shell parameters}
\noindent We start with the equation of the Gyroid surface
\[
f(\bar{x},\bar{y},\bar{z}) = \sin(\bar{x}) \cos(\bar{y}) + \sin(\bar{z}) \cos(\bar{x}) + \sin(\bar{y}) \cos(\bar{z}) = C, \tag{B1}
\]
where
\[
\bar{x} = \frac{\pi (x - a)}{a}, \quad 
\bar{y} = \frac{\pi (y - a)}{a}, \quad
\bar{z} = \frac{\pi (z - a)}{a},
\]
and
\[
(x, y, z) \in [-a, a]^3~.
\]
\\
We have to first find the location $y=y_0$ on the $x=a$ plane where the partial derivative wrt y vanishes.
Taking partial derivative on both sides of Eq. (B1) wrt y yields:

\begin{align*}
& \quad \quad ~~ \sin(\bar{x})(-\sin \bar{y})\frac{\pi}{a}
+ \cos(\bar{x})\cos(\bar{z})\frac{\pi}{a} \cancelto{0}{\frac{\partial z}{\partial y}}\\
& \quad \quad ~+ \sin(\bar{y})(-\sin \bar{z})\frac{\pi}{a}\cancelto{0}{\frac{\partial z}{\partial y}}
+ \cos(\bar{z})\cos(\bar{y})\frac{\pi}{a} = 0\\\\
&\Rightarrow \quad \sin(0)(-\sin(\bar{y}))\frac{\pi}{a}+\cos(\bar{z})\cos(\bar{y})\frac{\pi}{a} = 0\\
&\Rightarrow \quad \cos(\bar{z})\cos(\bar{y}) = 0\\
&\Rightarrow \quad \cos(\bar{y}) = 0
\end{align*}
Thus,
\[
\frac{\pi}{a}(y_0 - a) =-\frac{3\pi}{2} ,-\frac{\pi}{2}
\quad \Rightarrow \quad
y_0 = \pm \frac{a}{2}.
\tag{B2}
\]
\\
Now, solving for z\\
From Eq.(B1), for \( x = a \) and \( y = \frac{a}{2} \):
\begin{align*}
&\sin(\bar{z}) \cos(0) + \sin\left(-\frac{\pi}{2}\right)\cos(\bar{z}) = C, \\
&\Rightarrow \quad \sin(\bar{z}) - \cos(\bar{z}) = C\\
&\Rightarrow \quad \frac{1}{\sqrt{2}} \sin(\bar{z}) -\frac{1}{\sqrt{2}} \cos(\bar{z}) = \frac{C}{\sqrt{2}}\\
&\Rightarrow \quad \sin\left(\bar{z} - \frac{\pi}{4}\right) = \frac{C}{\sqrt{2}}\\
&\Rightarrow \quad \bar{z} = \frac{\pi}{4} + \sin^{-1}\!\left(\frac{C}{\sqrt{2}}\right)\\
&\Rightarrow \quad z = \frac{5a}{4} + \frac{a}{\pi} \sin^{-1}\!\left(\frac{C}{\sqrt{2}}\right).
\tag{B3}
\end{align*}
\\
Now, determining the constants $C_2$ and $C_3$

\begin{align*}
& \qquad ~~\frac{t}{2} 
= z(C_2) - z(C_1) \\
&\Rightarrow \quad \frac{t}{2}= \left(\frac{5a}{4} + \frac{a}{\pi}\sin^{-1}\!\frac{C_2}{\sqrt{2}}\right)
-\left(\frac{5a}{4} + \frac{a}{\pi}\sin^{-1}(0)\right)\\
&\Rightarrow \quad \frac{t}{2} = \frac{a}{\pi} \left[\sin^{-1}\!\left(\frac{C_2}{\sqrt{2}}\right)-\sin^{-1}(0) 
\right]\\
&\Rightarrow \quad\frac{n d_p}{2} = \frac{a}{\pi} \left[\sin^{-1}\!\left(\frac{C_2}{\sqrt{2}}\right)+\pi \right]\\
&\Rightarrow \quad C_2 = \sqrt{2} \sin\!\left(\left(\frac{n d_p}{2a} - 1\right)\pi\right)\\
&\Rightarrow \quad C_2 = -\sqrt{2} \sin\!\left(\pi - \frac{n \pi}{2}\frac{d_p}{a}\right)\\
& \therefore ~ \quad C_2 = -\sqrt{2} \sin\!\left(\frac{n \pi}{2}\frac{d_p}{a}\right)
\tag{B4}
\end{align*}
\\

Similarly,
\begin{align*}
& \frac{t}{2} 
= z(C_1) - z(C_3) \\
\therefore \quad &C_3 = \sqrt{2} \sin\!\left(\frac{n \pi}{2}\frac{d_p}{a}\right)
\tag{B5}    
\end{align*}

\end{document}